\documentclass{aa}
\usepackage[varg]{txfonts}
\usepackage[utf8]{inputenc}
\usepackage{graphicx}
\usepackage{array}
\usepackage{fp}
\usepackage{siunitx}
\usepackage{pdflscape}
\usepackage{numprint}
\usepackage{longtable}
\usepackage{multicol}
\usepackage{rotating}
\usepackage{microtype}
\usepackage{cuted}
\usepackage{flushend}
\usepackage{epstopdf}
\usepackage{supertabular}
\usepackage{afterpage}

\begin{document}

\title{Determination of spectroscopic parameters for 313 M dwarf stars from their APOGEE Data Release 16 H-band spectra}

\author{Pedro Sarmento \inst{1,2}
\and Bárbara Rojas-Ayala \inst{3}
\and Elisa Delgado Mena \inst{1}
\and Sergi Blanco-Cuaresma \inst{4}}

\institute{Instituto de Astrofísica e Ciências do Espaço, Universidade do Porto, CAUP, Rua das Estrelas, PT4150-762 Porto, Portugal e-mail:
pedro.sarmento@astro.up.pt
\and Departamento de Física e Astronomia, Faculdade de Ciências, Universidade do Porto, Portugal
\and Instituto de Alta Investigación, Universidad de Tarapacá, Casilla 7D, Arica, Chile
\and Harvard-Smithsonian Center for Astrophysics, 60 Garden Street, Cambridge, MA 02138, USA }

\abstract
{The scientific community's interest on the stellar parameters of M dwarfs has been increasing over the last few years, with potential applications ranging from galactic characterization to exoplanet detection.}
{The main motivation for this work is to develop an alternative and objective method to derive stellar parameters for M dwarfs using the H-band spectra provided by the Apache Point Observatory Galactic Evolution Experiment (APOGEE).}
{Synthetic spectra generated with \textit{iSpec}, \textit{Turbospectrum}, \textit{MARCS} models atmospheres and a custom made line list including over 1 000 000 water lines, are compared to APOGEE observations, and parameters are determined through $\chi^2$ minimization.}
{Spectroscopic parameters ($T_\mathrm{eff}$, $[M/H]$, $\log g$, $v_{mic}$) are presented for a sample of 313 M dwarfs, obtained from their APOGEE H-band spectra. The generated synthetic spectra reproduce observed spectra to a high accuracy level. The impact of the spectra normalization on the results are analyzed as well.}
{Our output parameters are compared with the ones obtained with APOGEE Stellar Parameter and Chemical Abundances Pipeline (ASPCAP) for the same stellar spectrum, and we find that the values agree within the expected uncertainties. Comparisons with other previous near-infrared and optical literature are also available, with median differences within our estimated uncertainties found in most cases. Possible reasons for these differences are explored. The full H-band line list, the line selection for the synthesis, and the synthesized spectra are available for download, as are the calculated stellar parameters.}

\keywords{stars: fundamental parameters - stars: M dwarfs - techniques: spectroscopy}

\titlerunning{Determination of M dwarf parameters}
\authorrunning{Pedro Sarmento et al.} 
\maketitle

\section{\label{intro} Introduction}

M dwarfs are main sequence stars with masses ranging from $0.6\,M_{\odot} \geq M_{*}\geq\,0.08\,M_{\odot}$, radii between $0.6\,R_{\odot}\geq R_{*}\geq\, 0.1\,R_{\odot}$, and $T_\mathrm{eff}$ from $3800\,K\geq T_\mathrm{eff}\geq\,2300\,K$ \citep{delfosse2000accurate}. They are the most abundant stars in the universe, with about 7 in every 10 stars in the Milky Way being M dwarfs, and accounting for most of its mass \citep{henry2006solar}. M dwarfs are the faintest main-sequence stars, with the largest and brightest of them having only $0.1\,L_{\odot}$. However, and despite their relative dimness, M dwarfs can be the key to a bright future in areas of astronomy such as galactic chemical evolution and exoplanet science. 

As M dwarfs have a longer lifespan than other larger and brighter main-sequence stars and they represent such a large fraction of existing stars in our galaxy, M dwarf characterization is essential for our understanding of the galactic history. Some examples are \cite{lepine2007revised}, which uses spectroscopic indices and the kinematic measurements of both disk dwarfs and halo subdwarfs to revise metallicity classes for these objects, and \cite{bochanski2007exploring}, that uses a spectroscopic sample of low-mass stars to investigate the properties of the thin and thick disks.

M dwarfs are very good targets for surveys dedicated to the discovery of Earth-like planets. Due to their faintness, their habitable zone is much closer to the star than it is for solar-type stars, and, due to their small radius and mass, the relative difference in size between any potential exoplanet and the host star is much smaller than for more massive stars. As the two most successful current methods for planet detection, radial velocity and transit, are indirect methods, this means that detecting exoplanets in the star's habitable zone is orders of magnitude easier for M dwarfs than for solar-type stars. Several programs therefore focus primarily on finding potentially habitable planets around M dwarfs, such as MEarths \citep{irwin2014mearth}, and CARMENES \citep{alonso2015carmenes} or HARPS \citep{bonfils2013harps}, among others, have also targeted M dwarfs with the purpose of planet detection. There have also been many studies detailing how life might be possible in planets around M dwarfs \citep[e.g.][]{segura2005biosignatures,scalo2007m,france2013ultraviolet}.

Stellar characterization through spectroscopy can provide the scientific community with fundamental stellar parameters such as effective temperature, metallicity, and surface gravity. These stellar parameters can be used, for example, to discover more information about the history of our galaxy through the study of metal-poor stars \citep{suda2008stellar}, improve our understanding of giant stars \citep{ness2016spectroscopic} or for the characterization of exoplanet host stars, and, indirectly, of exoplanets \citep[e.g,][]{Sousa2011a,santos2013sweet}.

Spectroscopic analysis of M dwarfs in the NIR has been a developing topic over the last few years, with works in lower resolution such as \cite{rojas2010metal,rojas2012metallicity}, a characterization of 133 M dwarfs using K-band (2.0-2.4\,micron, $S/N > 200$, $R\sim2700$) spectra, as well as \cite{mann2013prospecting}'s work with $1300<R<2000$ optical and infrared spectra, and \cite{terrien2015near}'s catalog of 886 M Dwarfs in the full NIR (0.8-2.4 micron, $R\sim2000$). More recently, works have moved to higher resolution, with publications such as \cite{veyette2017physically} providing an analysis of 29 M dwarfs from their NIRSPEC (Keck II) Y-band ($\sim 1\mu m $, $R\sim 25\,000$) spectra with the help of PHOENIX models. The high-resolution studies of \cite{onehag2012m}, using CRIRES $R\sim50 000$ J band spectra, as well as the more recent \cite{passegger2019carmenes}'s characterization of 300 stars by fitting PHOENIX models to their high-resolution ($R=80\,000-100\,000$) CARMENES optical and near-infrared spectra expanded our understanding of the spectra of these stars. 

Our work positions itself as a continuation and extension of these works, providing a new pipeline for spectroscopic parameter derivation from the Apache Point Observatory Galactic Evolution Experiment \citep[APOGEE,][]{majewski2017apache}'s mid-high resolution ($R\sim 22\,000$) H-band (1.5-1.7\,micron) stellar spectra.

This paper is structured as follows: Section \ref{Data} contains a description of both APOGEE and our sample stars, detailing available previous literature analysis of them; Section \ref{method} details the methodology used for our characterization of stellar spectra ; Section \ref{Results} contains our main results, including both an H-R diagram of our 313 characterized stars and an example of a synthesized stellar spectrum; Section \ref{Discussion} includes our analysis of the derived parameters, including literature comparisons; and Section \ref{Conclusions} details our main conclusions and future possibilities for the pipeline.

This work comes as a follow-up to \cite{sarmento2020derivation}, using similar methods and expanding the parameter space of its analysis into late-K and early-M dwarf stars.

\section{Data \label{Data}}

All M dwarf spectra used for stellar characterization in this paper comes from the APOGEE survey. APOGEE is an H-band (1.5-1.7 micron) Sloan Digital Sky Survey program that focuses on obtaining $R \sim 22\,500$ stellar spectra with a 300-fiber spectrograph. It is split between APOGEE-N, using the Sloan 2.5\,m telescope at the Apache Point Observatory in New Mexico \citep{gunn20062}, and APOGEE-S, which uses the 2.5\,m duPont telescope at the Las Campanas Observatory in Chile \citep{bowen1973optical}. It targets mostly red giants and provides public spectra for more than 200\,000 stars in its Data Release 14 \citep[DR14,][]{holtzman2018apogee}. APOGEE has observed FGK and M dwarfs for calibration purposes or as part of ancillary programs \citep{zasowski2013target}. 

Their spectroscopic parameters (effective temperature, surface gravity, microturbulence, macroturbulence, rotation, overall metal abundance $[M/H]$ , relative $\alpha$-element abundance $[\alpha/M]$ (determined by fitting simultaneously lines of O, Mg, Si, S, Ca, and Ti), carbon $[C/M]$, and nitrogen $[N/M]$ abundances) have been derived with APOGEE Stellar Parameter and Chemical Abundances Pipeline \citep[ASPCAP,][]{perez2016aspcap}. ASPCAP also determines abundance for 24 different elements. APOGEE's Data Release 16 \citep[DR16,][]{APOGEE_Dr16, 2020ApJS..249....3A,2020AJ....160..120J} is the most recent release of APOGEE data, and it is the first one to include results obtained with APOGEE-2. The spectra for more than 430 000 stars is included in this data release. The previously released data was analyzed again, with small changes to the data processing and ASPCAP analysis procedures. Comparing with previous data releases, the most important of these changes is an expansion of the parameter space of the characterized stars, with published calibrated $\log g$ values up to $6.0\,dex$ rather than $4.5\,dex$, and $T_\mathrm{eff}$ down to $3200\,K$. ASPCAP works by searching and interpolating a grid of synthetic spectra to find the best match for each observed spectrum, adopting the parameters of the synthetic spectrum as the preliminary parameters for each star. These parameters are then calibrated to follow the theoretical models, with the calibrated values being the final parameters for each star.

In addition to the spectra characterized by ASPCAP DR16, we decided to apply our method to the spectra of the same stars, as published in APOGEE DR14 \citep{holtzman2018apogee}. These results are included in the paper for reference and comparison between the two data releases, and because they were the main results of the first author's thesis. The main differences between the two data releases, in addition to small cleanups of the spectra, including the removal of cosmic rays, are the parameter calibrations, which, for $\log g$ and $T_\mathrm{eff}$, are now extended to better characterize the M dwarf parameter range.

\subsection{Sample selection}

All stars in our sample were chosen as part of APOGEE's ancillary M dwarf program. The goal of this program was to constraint the rotational velocities and compositions of over 1400 M dwarfs and to detect their low mass companions through RV variability, as published in \cite{deshpande2013sdss}. The targets are drawn primarily from two sources, the LSPM-North catalog of nearby stars \citep{lepine2005catalog} and the catalog of nearby M dwarfs by \cite{lepine2011all}. In order to obtain their final selection, magnitude and color cuts ($7 <  H < 12\,; V-K >5.0 \,; 0.4 <J-H<0.65\,; 0.1<H-Ks<0.42$) were applied to stars from those catalogs that fell on APOGEE's observation fields. In addition to both of these proper-motion selected catalogs, additional known planet hosts and other stars with available rotational speed, metallicity, or radial velocity estimates were also included for calibration purposes. Although \cite{deshpande2013sdss} published rotational velocities for these M dwarfs, they were unable to provide stellar parameters for all stars in the sample due to the inability to model the strong $H_{2}O$ bands found in the spectra. 

The sample analyzed in this work was filtered by $S/N>200$ to remove the spectra with worst quality from the sample. We chose this value as a limit because it both provided us with enough stars to perform a statistical analysis and was not so large that the time for the computational analysis became unpractical. The distribution of APOGEE's reported $S/N$ for the stars in our final sample is available in the bottom part of Fig. \ref{Histograms_Mdwarfs}. In addition to our star selection by $S/N$, 21 additional stars characterized by \cite{Souto2020} and 9 stars with confirmed exoplanets were included in the sample, despite having a lower $S/N$ ratio than the cutoff for the rest of our analyzed stars. In order to test the lower bounds of our method, 7 M dwarfs with ASPCAP $T_\mathrm{eff} < 3000\,$K were included in the sample as well. Our final M dwarf sample contains 313 stars. 

\subsection{Sample characterization}

\begin{figure}
\resizebox{\hsize}{!}{\includegraphics{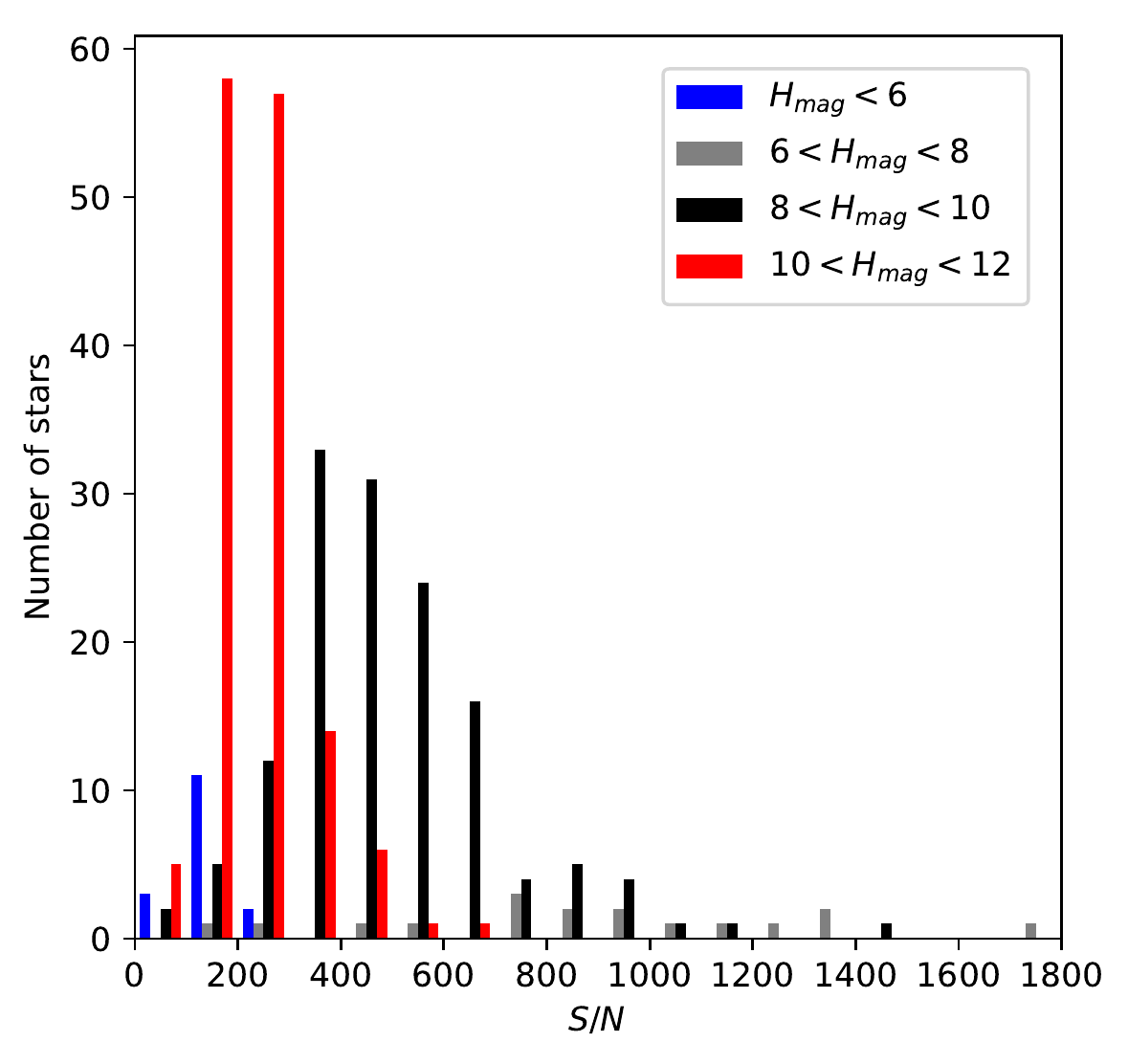}}
\caption{Histogram presenting both the H magnitude and the $S/N$ (from APOGEE) of our M dwarf sample stars.}
\label{Histograms_Mdwarfs}
\end{figure}

We display both the magnitude distribution and the APOGEE-estimated $S/N$ of our 313 sample stars in Fig. \ref{Histograms_Mdwarfs}. The figure shows how most of our sample has high $S/N$ spectra, and also shows how fainter stars tend to have lower $S/N$ as expected. The small clump of lower magnitude (4-6), $S/N<300$ stars corresponds to the M dwarfs from \cite{Souto2020} that have been characterized with interferometry.

\begin{figure}
\resizebox{\hsize}{!}{\includegraphics{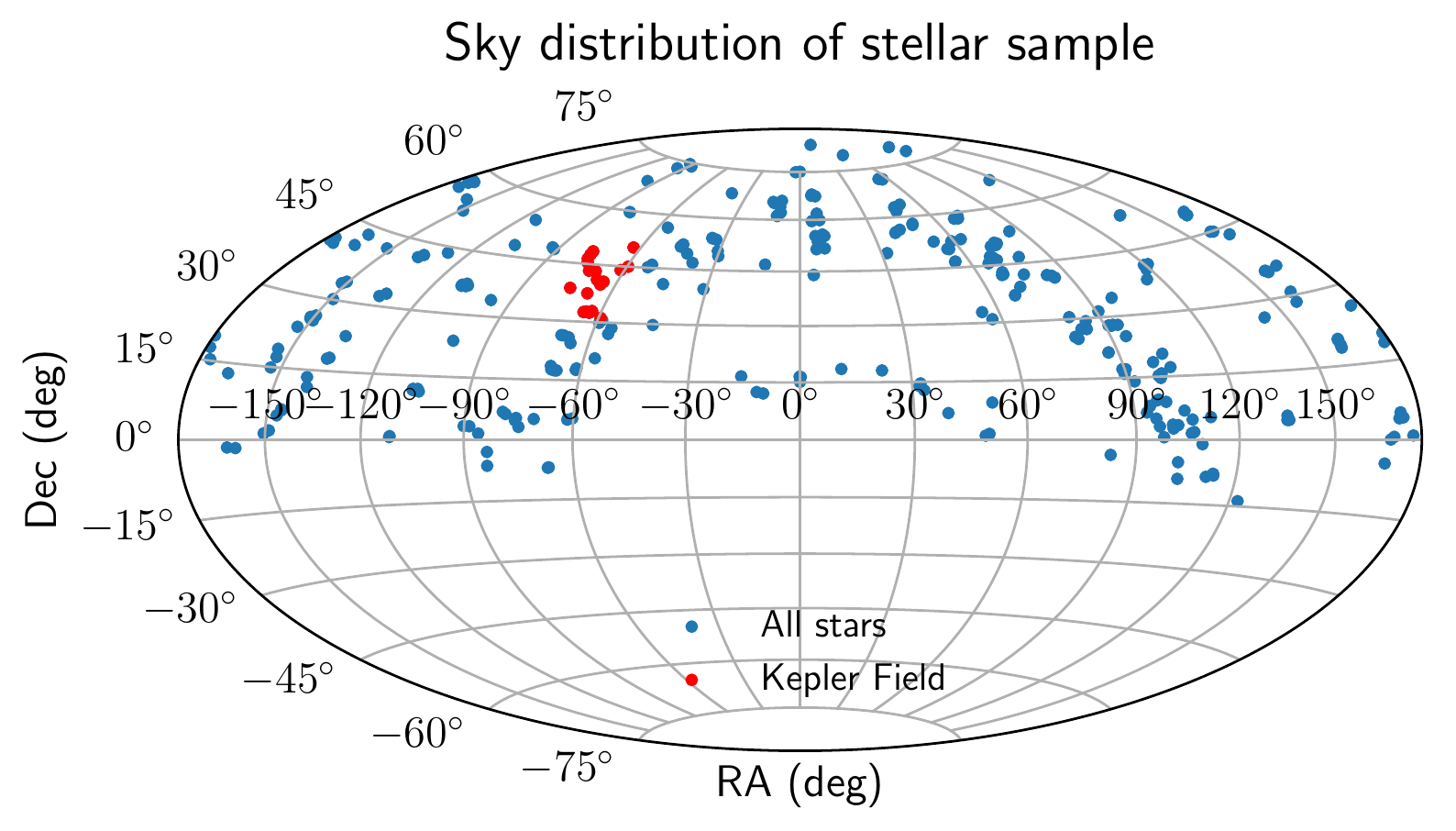}}
\caption{Map showing the location in the sky of our sample M dwarfs. Kepler target field is highlighted in red.}
\label{declination_Mdwarfs}
\end{figure}

We display the location in the sky of all M dwarfs in the sample, showing both their Right Ascension (RA) and Declination (Dec) in Fig. \ref{declination_Mdwarfs}. Most of the stars are located in the northern hemisphere (Dec > 0\,deg), as they were observed by APOGEE-2N. Nineteen stars in our sample are located in the Kepler field \citep{latham2005kepler}, fourteen of which have been observed by the Kepler Space Telescope.


\subsection{Available literature values for the sample\label{SampleParams}}

We cross-matched our M dwarf sample with different surveys and works related to spectroscopic stellar characterization. We found 8 works with stars in common with our sample that serve as our literature comparisons - ASPCAP \citep{perez2016aspcap}, \cite{terrien2015near}, \cite{gaidos2014trumpeting}, \cite{Hejazi2019Chemical}, \cite{Souto2020}, \cite{passegger2019carmenes}, \cite{rajpurohit2018carmenes} and \cite{rajpurohit2018apogee}. This subsection is dedicated to explanations of these works and their determined parameters for the stars in common between them and our M dwarf sample.

\subsubsection{ASPCAP}

ASPCAP \citep{perez2016aspcap}, as the pipeline dedicated to derivation of spectroscopic parameters for stars observed with APOGEE, published their estimates for stars in our sample as well. The ASPCAP pipeline provides both a preliminary set of output parameters for each observed star as well as a final set of calibrated parameters focused on its main target of giant stars. However, their calibrated parameter ranges do not include all M dwarfs in our sample, as their lower boundary for $T_\mathrm{eff}$ is $3200$\,K. Their values can still be an important reference for our pipeline, as they have been shown to be consistent by \cite{schmidt2016examining}, which performed an extensive analysis on both the $T_\mathrm{eff}$ and $[M/H]$ values ASPCAP produced for 3834 M dwarfs, comparing them to photometric and interferometric parameters for the same stars. Based on interferometric data from \cite{Boayajian2013}, color-$T_\mathrm{eff}$ relations from \cite{mann2015constrain}, and the infrared flux technique published by \cite{casagrande2008m}, they concluded that the ASPCAP $T_\mathrm{eff}$ is accurate to about $100\,\mathrm{K}$ for stars between 3550–4200\,K, and $[M/H]$ values are accurate to about $0.18\,\mathrm{dex}$. This means that, despite ASPCAP's main focus being on giant stars and not M dwarfs, we can take their measurements of $T_\mathrm{eff}$ and $[M/H]$ as consistent for the parameters of M dwarfs in our sample, and we will thus use them as a comparison value for our analysis.

\begin{figure}
\resizebox{\hsize}{!}{\includegraphics{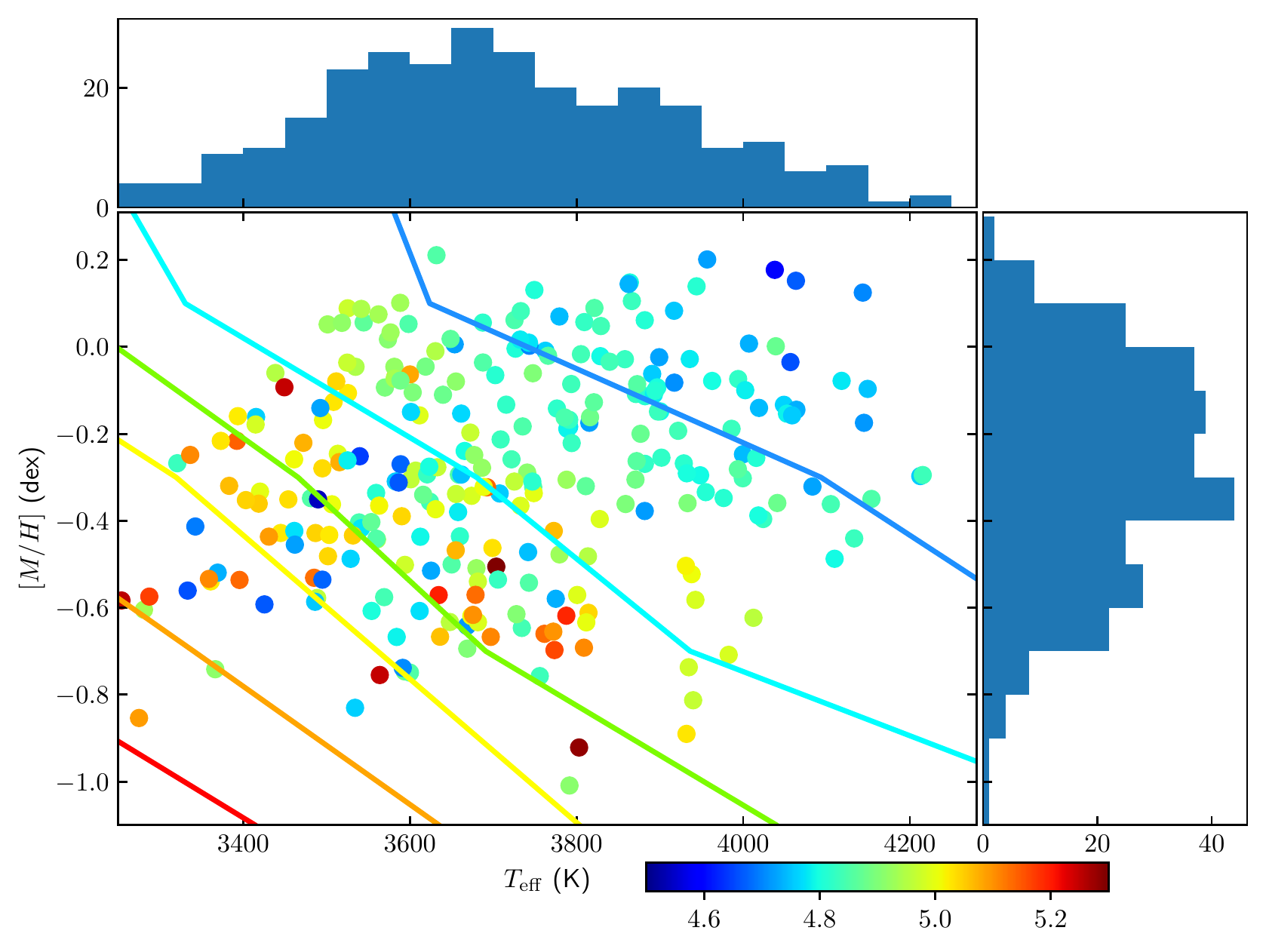}}
\caption{ASPCAP $T_\mathrm{eff}$ and $[M/H]$ values for M dwarfs in sample, with overplotted PARSEC isochrones \citep{bressan2012parsec} for an age of 5Gyr created with different $\log g$ values (scale is in dex). Points are color-coded based on the metallicity value derived for each star.}
\label{Mdwarfs_ASPCAP_params}
\end{figure}

From the 313 stars in our sample, ASPCAP has published their final, calibrated $[M/H]$, $T_\mathrm{eff}$, and $\log g$ for 283 stars. These values are displayed in Fig. \ref{Mdwarfs_ASPCAP_params}, with the color indicating the ASPCAP $\log g$ values for each star. The figures shows that their $T_\mathrm{eff}$ values are concentrated around 3600-3800\,K, and $[M/H]$ are between -0.4 and +0.0\,dex. The metallicity distribution follows an approximate Gaussian shape, with an average value of -0.29\,dex and a standard deviation of 0.27\,dex. These results show that a significant fraction of stars in our sample are early M dwarfs. According to ASPCAP results, our sample is, on average, more metal-poor than the solar neighborhood, which has mean metallicity values around -0.20\,dex \citep{holmberg2007geneva}.


\begin{figure}
\resizebox{\hsize}{!}{\includegraphics{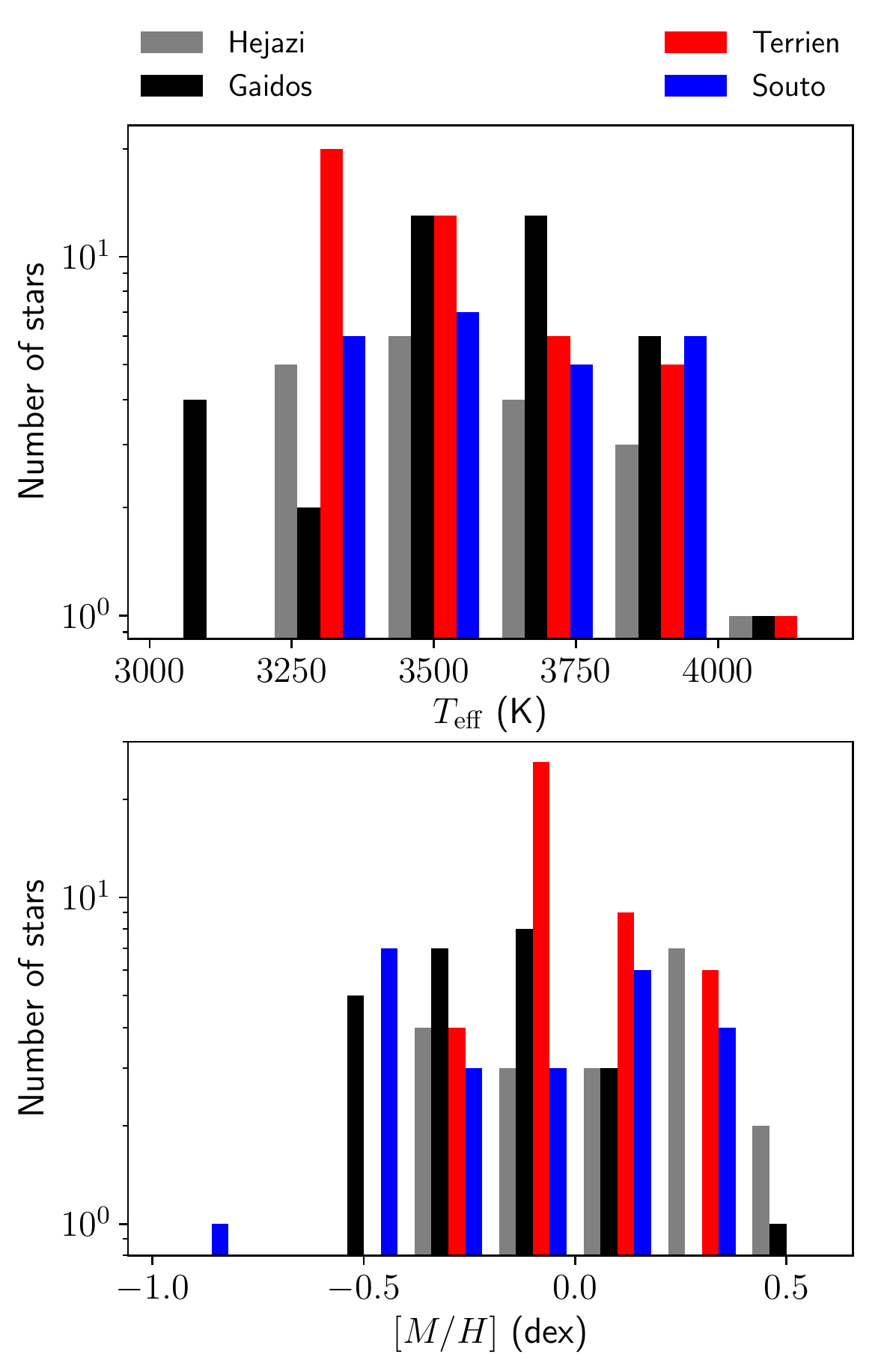}}
\caption{$T_\mathrm{eff}$ (TOP, a) and $[M/H]$ (BOTTOM, b) values from 4 literature sources: \cite{gaidos2014trumpeting} (black, $T_\mathrm{eff}$ for 40 stars, $[Fe/H]$ for 24 stars), \cite{Hejazi2019Chemical} (orange, filled bars, 19 stars), \cite{terrien2015near} (red, 45 stars), and an aggregate between \cite{souto2017}, \cite{souto2018stellar}, and \cite{Souto2020} (blue, 24 stars). Number of stars is displayed with logarithmic notation for visibility.}
\label{3sources}
\end{figure}

\subsubsection{\cite{terrien2015near}}

From the 886 stars characterized in \cite{terrien2015near}, 45 are included in our sample. Their spectra have wide coverage ($0.8-2.4\,\mathrm{\mu m}$) but low resolution ($R\sim2000$). Their $T_\mathrm{eff}$ values are determined by analysis of water indices in the K band and have uncertainties above 100\,K, and their $[M/H]$ values are measured using empirical spectroscopical calibrations with uncertainties around 0.12\,dex. Their published parameters are shown in Fig. \ref{3sources}. Comparing their parameter distribution with the full ASPCAP sample, we find that \cite{terrien2015near} is more representative of later-type M dwarfs, having a significant fraction of stars with published $T_\mathrm{eff}$ between 3200-3500\,K.

\subsubsection{\cite{souto2017}, \cite{souto2018stellar}, \cite{Souto2020}}

We find that some of the best characterized stars in the APOGEE sample are the M dwarfs that were analyzed by \cite{souto2017}, \cite{souto2018stellar}, and \cite{Souto2020}. That is because these works use spectral synthesis and a combination of \textit{MARCS} and \textit{PHOENIX} model atmospheres to determine both stellar parameters and chemical abundances for 8 different elements using APOGEE spectra. Their reported uncertainties are $T_\mathrm{eff}\pm100\,$K, $\log g\pm0.2\,$dex and $[Fe/H]\pm0.1\,$dex. A total of 24 different M dwarfs were characterized by this method - Kepler 138 and Kepler 186 in \cite{souto2017}, Ross 128 in \cite{souto2018stellar}, and 21 other stars in \cite{Souto2020}. Among all the stars in our sample previously characterized in the literature, the most metal-poor one is 2M03150093+0103083, with $[Fe/H]=-0.91\,dex$, and it was characterized by \cite{Souto2020}.

\subsubsection{\cite{gaidos2014trumpeting}}

\cite{gaidos2014trumpeting} analyzed optical spectra of nearby K and M dwarfs with $R\sim1000$ and determined their spectroscopic parameters by fitting model spectra to observations. Their estimated errors vary between stars, but are around 100\,K for $T_\mathrm{eff}$ and 0.12\,dex for $[Fe/H]$.. From the 2970 stars with $d<50\,pc$ observed by \cite{gaidos2014trumpeting}, 28 are included in our sample, and their parameters are shown in Fig. \ref{3sources}. The reduced number of stars, combined with the lack of $[Fe/H]$ available for half the stars in the sample, does not allow for an extensive statistical analysis of this stellar subsample. Nevertheless, it is possible to notice \cite{gaidos2014trumpeting} reported $T_\mathrm{eff}$ values between 3100-3500\,K for a greater number of stars in this subsample than in the main ASPCAP one, helping us characterize the lower $T_\mathrm{eff}$ stars in our sample. \cite{gaidos2014trumpeting} also found metallicity for stars in our sample across a larger parameter space ($-0.6<[Fe/H]<+0.4\,$dex) than ASPCAP.

\subsubsection{\cite{Hejazi2019Chemical}}

\cite{Hejazi2019Chemical} published chemical properties for 1544 high proper-motion M dwarfs and subdwarfs from low-mid resolution ($\sim 2000-4000$) optical spectra. A template-fit method was developed, based on the measurement of TiO and CaH molecular bands near $7000 \AA$. The analysis of 48 binary systems suggests precision levels of $\pm 0.22$ for $[M/H]$, $\pm 0.08$ for $[\alpha/Fe]$, and  $\pm 0.16$ for the combined index $[M/H] + [\alpha/Fe]$. We find 19 stars in common between our observed sample and their study. From Fig. \ref{3sources}, we can verify that the characterized stars are dispersed across a wide range of temperatures, with a stronger concentration of stars around $T_\mathrm{eff} = 3600\,K$. As for the metallicities, we find stars both above and below solar neighborhood values, but no star within $-0.1<[M/H]<+0.1\,$dex, which must be considered when comparing our results to theirs.


\subsubsection{\label{passegger}\cite{passegger2019carmenes}}

We cross-matched our sample with the 300 stars characterized by \cite{passegger2019carmenes} and found 14 stars in common. As mentioned in Section \ref{intro}, this work derived parameters for M dwarfs by using a $\chi^2$ to fit \textit{PHOENIX-SESAM} models to high-resolution ($R\sim90\,000$) CARMENES spectra in both the visible and infrared wavelength ranges. Their reported uncertainties depend on each star's rotational velocity and, for NIR spectra, are as low as 51\,K for $T_\mathrm{eff}$, 0.07 for $\log g$, and 0.16 for $[Fe/H]$.

\subsubsection{\cite{rajpurohit2018carmenes}}

Another study with CARMENES spectra is \cite{rajpurohit2018carmenes}. This work focused on matching BT-Settl model spectra to CARMENES optical and near-infrared observations of 292 M dwarf spectra. Since this work determines parameters by matching observed spectra to a grid of previously computed models, the reported uncertainties correspond to the size of the grid, 100\,K for $T_\mathrm{eff}$ and 0.1\,dex for both $\log g$ and $[M/H]$. We find 12 stars in common between our sample and the one characterized by \cite{rajpurohit2018carmenes}, which are part of the 14 stars later characterized by \cite{passegger2019carmenes} and listed in section \ref{passegger}.

\subsubsection{\cite{rajpurohit2018apogee}}

The same technique was applied to 45 M dwarfs observed with APOGEE in \cite{rajpurohit2018apogee}. We have 8 stars in common, with six of them having $T_\mathrm{eff}<3300\,$K and are some of the coldest stars in our sample. The reported uncertainties for the parameters of these stars are $T_\mathrm{eff}\pm100\,$K, $\log g\pm0.3-0.5\,$dex and $[M/H]\pm0.05\,$dex. A comparison between our derived parameters and the previously available parameters for these stars is available in Fig.\ref{Raj_Pass_comp}.

We found \cite{lopez2019effective} as an example of another large scale, high resolution study of spectroscopical parameters for M dwarfs in the infrared, but, unfortunately, it shares no stars in common with APOGEE. The best spectroscopic parameters available in literature for our sample stars are the ones cited in this section and ASPCAP values. Therefore, comparisons between our results and literature values are made against the works mentioned above.

\section{\label{method} Methodology}

This section includes the steps in our method that are specific for M dwarfs. The full rundown of our pipeline is available at \cite{sarmento2020derivation}. Summarizing, our first step is to normalize the APOGEE observed spectra for each star using the template method and a synthetic spectrum. Then, we created custom software that uses \textit{iSpec} \citep{blanco2014determining,ispec2019sbc} and \textit{Turbospectrum} \citep{plez1998,plez2012turbospectrum} to derive the atmospheric parameters through a $\chi^2$ minimization algorithm that compares observed and synthetic spectra. These synthetic spectra are created using \textit{MARCS} \citep{MARCS} stellar atmospheric models and a custom line list. The best fitting synthetic spectra are selected through $\chi^2$ minimization, using MPFIT \citep{markwardt2009non} internally through \textit{iSpec}. Finally, the parameters of the best matching synthetic spectra are taken as our measurements for the stellar parameters of each star.

We created routines for the download, format conversion, normalization of the sample spectra, and have made the pipeline as automated as possible, but the $\chi^2$ minimization and the synthesis of the best fitting spectra are done with \textit{Turbospectrum}, controlled by \textit{iSpec} python code. The custom routines are programmed in Python 3 and for the 2020 distribution of iSpec. They are included in a public repository together with the full results, line lists, line masks, and everything required to replicate the process described in this paper.

\subsection{\label{lineMdwarf} Line list}

The line list used for M dwarf spectral syntheses is more complex than the one used for the spectra of FGK stars, as these stars have many water ($\textrm{H}_{2}\textrm{O}$) lines that are not visible in hotter stars. These water lines form a thick blanket that makes the flux continuum harder to identify, as shown in figure \ref{teff_comp}. We use the water line list first published in \cite{barber2006high}, retrieved from Bertrand Plez's personal website \footnote{http://www.pages-perso-bertrand-plez.univ-montp2.fr/}. We note here that a more recent work on water lines has been published by \cite{polyansky2018exomol}, but preference was given to usage of the water line list from \cite{barber2006high} as it was already formatted for Turbospectrum and of a more computationally manageable size. A total of 1 263 825 water lines were included in the M dwarf line list, along with the 85334 lines from other molecules and atoms already available at \cite{sarmento2020derivation}, made from a combination of the Vienna Atomic Line Database \citep[VALD,][]{piskunov1995vald} and the APOGEE line list \citep{shetrone2015sdss}. This resulted in a final list of 1 349 159 lines.

\subsection{\label{MdwarfNorm} Spectra Normalization}

\begin{figure}
\resizebox{\hsize}{!}{\includegraphics{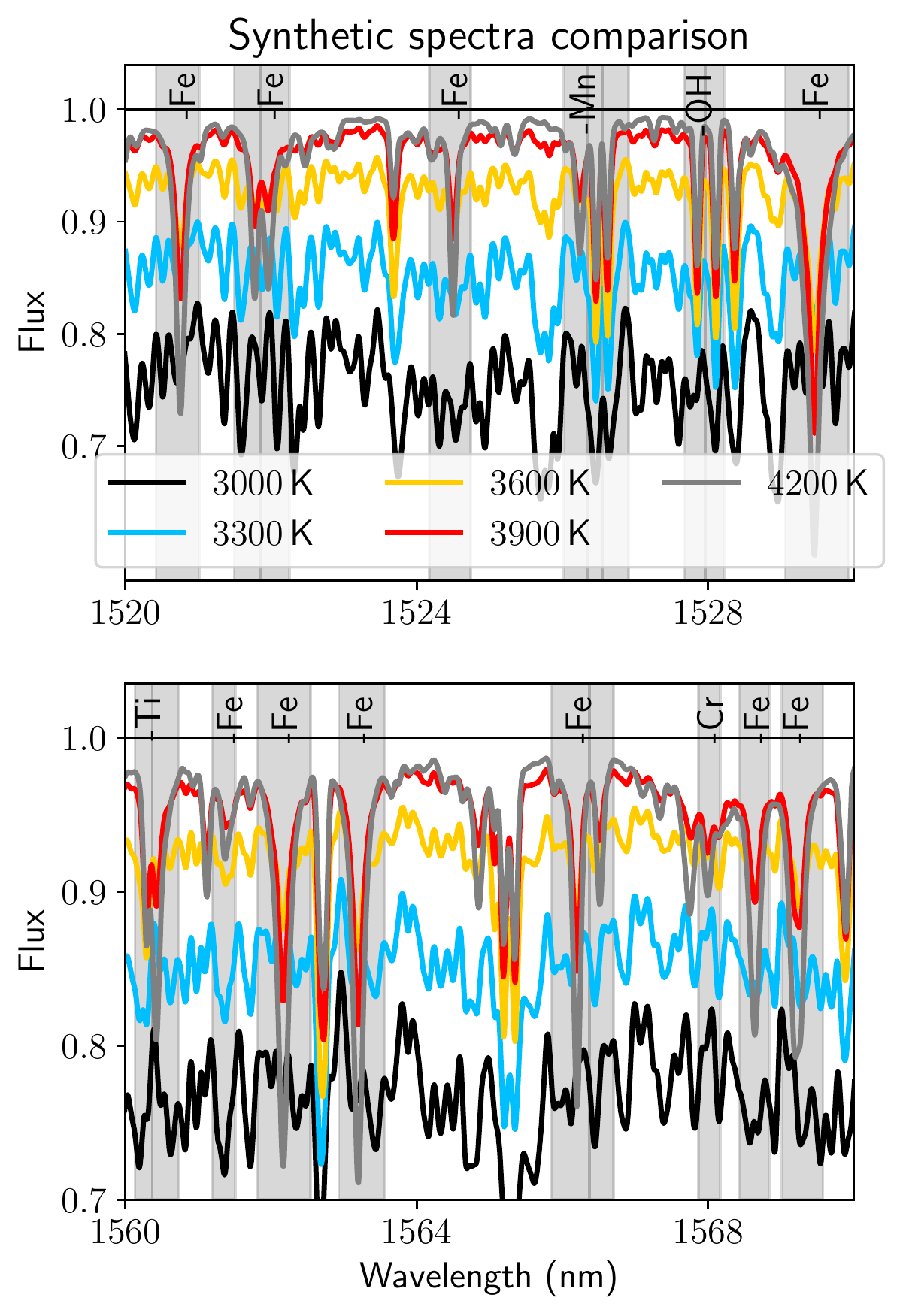}}
\caption{Comparison between synthetic spectra made with different $T_\mathrm{eff}$. The spectra corresponds to synthesis made with $T_\mathrm{eff}$ 4200K (gray), 3900K (red), 3600K (orange), 3300K (orange) and 3000K (black).}
\label{teff_comp}
\end{figure}

As we can see in Fig.\ref{teff_comp}, a $300\,\mathrm{K}$ difference in $T_\mathrm{eff}$ can mean a 0.05 difference in the continuum flux for a star, which is enough to turn an accurate fit into a terrible one. Therefore, in order to accurately normalize an M dwarf's H-band spectrum, an accurate initial input stellar $T_\mathrm{eff}$ for that star is required. Normalizations with templates synthesized with $T_\mathrm{eff}$ closer to the star's real $T_\mathrm{eff}$ will result in continuum values closer to the real ones, meaning the synthetic spectra will closely resemble the observed ones. Therefore, the first important step towards deriving a star's atmospheric parameters is to have good estimates for their $T_\mathrm{eff}$ and metallicity.

Given the lack of a consistent literature source for the $T_\mathrm{eff}$ of our sample stars, and in order to minimize the effect of the visual choice of $T_\mathrm{eff}$ based on the normalized spectra, the spectrum of each star in our sample was normalized using synthetic spectra with 148 different combinations of $T_\mathrm{eff}$, $\log g$ and $[M/H]$, corresponding to values taken from PARSEC isochrones \citep{bressan2012parsec} for stars with ages between $10^9$ and $10^{10}$ years. This approach is done to avoid unrealistic combinations of output parameters. We selected isochrones with a 0.2\,dex spacing in $[M/H]$ values, rounding $T_\mathrm{eff}$ values to the nearest multiple of 100\,K, and $\log g$ being rounded to the nearest multiple of 0.1\,dex. These values are selected to both have an amount of normalizations that can be computationally generated in a short amount of time and to have an even distribution across our expected parameter space for M dwarfs, as our sample stars are expected to have parameters approximating one of these combinations. The full list of parameter combinations is available at appendix \ref{table:parameter_normalization}.

\begin{figure}
\resizebox{\hsize}{!}{\includegraphics{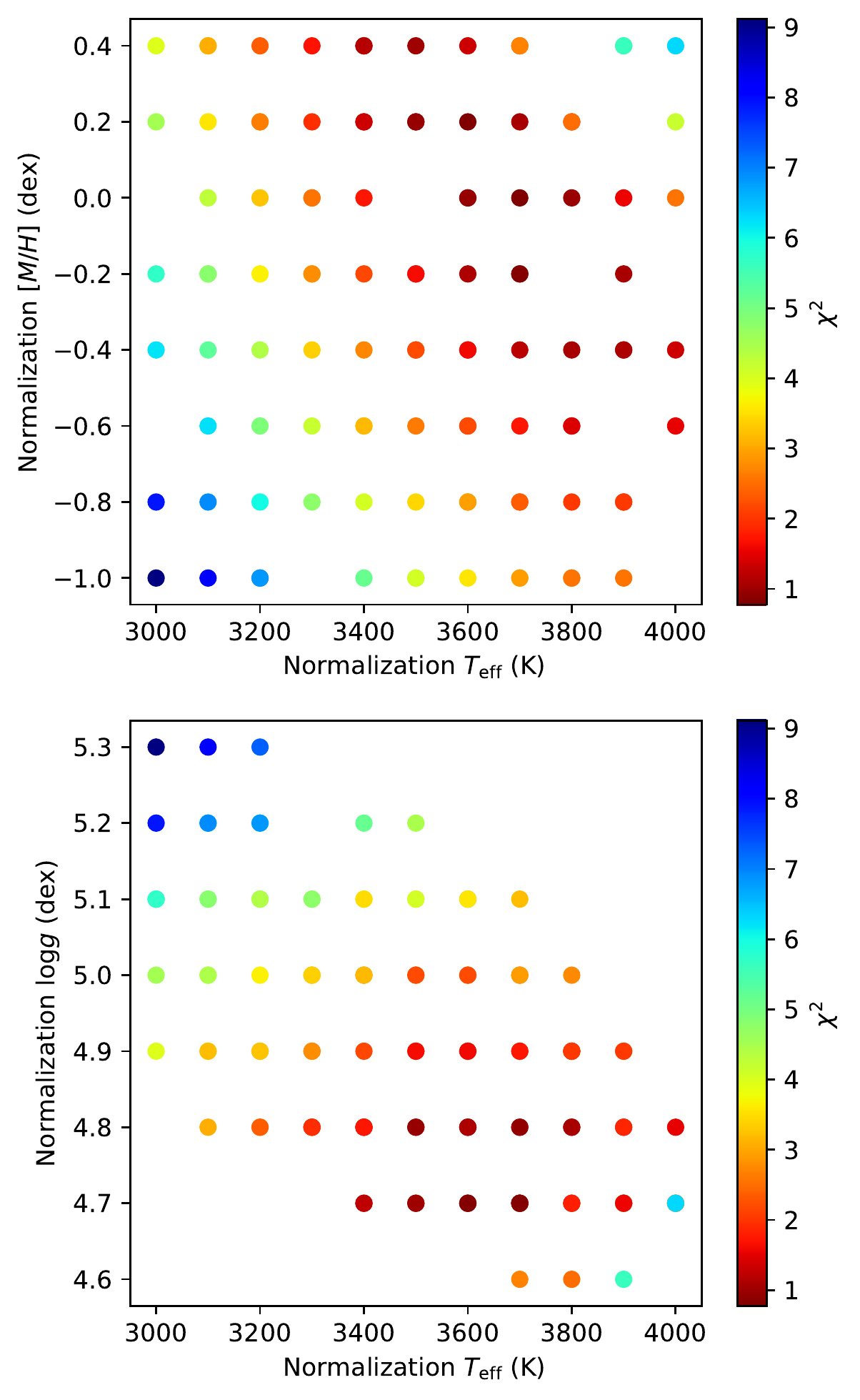}}
\caption{TOP (a): Scatter plot displaying the $\chi ^2 $ values measured for different $T_\mathrm{eff}$ and metallicity combinations for templates for the normalization of star 2M05201152+2457212. BOTTOM (b): Similar plot with $T_\mathrm{eff}$ and $\log g$ for the same star.}
\label{norm_chi2}
\end{figure}

Afterwards, we compared each of the normalized spectra to the template used to create the normalization. By calculating and minimizing the $\chi ^2$ across our line mask region (see section \ref{linemaskMd}) between the observed and the template spectra, the parameters for the best normalization template are defined. Fig. \ref{norm_chi2} displays the $\chi ^2$ values for different normalization parameters for star 2M05201152+2457212. That figure shows how the $\chi ^2$ values depend strongly on the normalization template, and can indicate the quality of the template choice for normalization. It is also clear that templates with similar parameters have similar $\chi ^2$ values. This fact shows that these values are close to the real stellar parameters, and the normalization with a template with $T_\mathrm{eff}=3700\,\mathrm{K}$,$\log g=4.7\,\mathrm{dex}$, $[M/H] = 0.0\,\mathrm{dex}$ can be used for this star in the next steps of the pipeline.

\subsection{\label{linemaskMd} Line mask}

\begin{landscape}

\begin{figure}
\resizebox{\hsize}{!}{\includegraphics{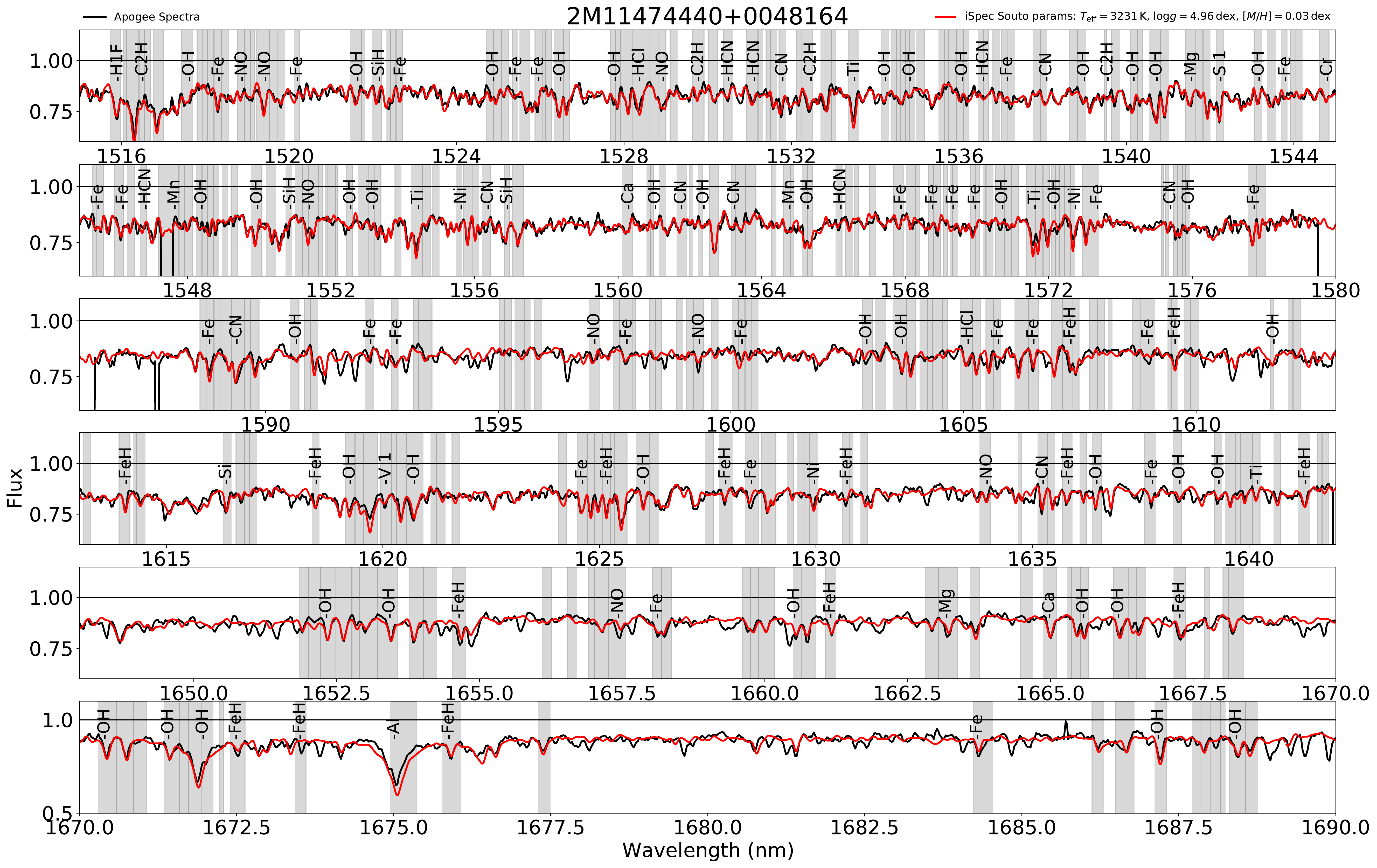}}
\caption{Comparison between the observed spectrum of Ross 128 (2MASS J11474440+0048164), normalized using the template method (black), and a synthetic spectrum created using the literature \citep{souto2018stellar} parameters for that star (red). The parameters used to generate the synthetic spectrum were $T_\mathrm{eff}=3231\,\mathrm{K}$,$\log g=4.96\,\mathrm{dex}$, $[M/H] = +0.03\,\mathrm{dex}$. The ASPCAP preliminary parameters for this star are $T_\mathrm{eff}=3212\,\mathrm{K}$,$\log g=4.45\,\mathrm{dex}$, $[M/H] = -0.60\,\mathrm{dex}$. The areas in gray represent the ones included in the line mask created from these spectra.}
\label{MdwarfMask}
\end{figure}

\end{landscape}

As explained in \cite{sarmento2020derivation}, a line mask instructs \textit{iSpec} on which spectral regions should be considered for comparing observed and synthetic spectra. A line mask specifically made for M dwarf stars was therefore required for the code to match an appropriate model to the observations. The star Ross128 (2M11474440+0048164), as an APOGEE M-dwarf with $S/N\sim229$ that has previously been characterized by \cite{souto2018stellar}, was chosen as the basis for this line mask. A synthetic spectrum was generated using the parameters found for the star by \cite{souto2018stellar} ($T_\mathrm{eff} = 3231\,\mathrm{K}$, $[Fe/H] = 0.03\,\mathrm{dex}$ and $\log g=4.96\,\mathrm{dex}$). This synthetic spectrum was used to normalize the observed one and then compare it. Regions and spectral lines with a good agreement between both spectra were selected for the line mask displayed in Fig.\ref{MdwarfMask}. The areas where the synthesis did not reproduce well the observed spectrum were discarded to create the final line mask shown. We note that, despite the slight errors in the Al lines around 1675\,nm, we included these lines in our line mask. This was done because the two Al lines are the strongest lines we find across our wavelength range, and they are very well reproduced across other stars (see for example the appendix for Figs.\ref{Mdwarf_1}, \ref{Mdwarf_2}, \ref{Mdwarf_5}).

\subsection{Free parameters}

The free parameters used for the syntheses of M dwarf spectra are the same as the ones used in \cite{sarmento2020derivation} for the syntheses of the spectra of FGK stars - effective temperature ($T_\mathrm{eff}$), surface gravity ($\log g$), metallicity ($[M/H]$), micro-turbulent velocity ($v_{mic}$), projected rotational velocity ($v \sin i$), and spectral resolution. The input values used are, for $T_\mathrm{eff}$, $\log g$, and $[M/H]$, the ones used to create the template spectra for the normalization (see Section \ref{MdwarfNorm}). The initial values for the remaining free parameters are 1.06\,km/s for the $v_{mic}$, 1.6\,km/s for $v \sin i$, and 22\,000 for the resolution. We note that, despite leaving the resolution as a free parameter, our output values for this parameter are close to 22\,000. 

However, an important change required to keep the output parameters limited to realistic values is the restriction of possible output parameters to a given parameter space. Restricting the parameters to a given interval based on the starting values allows the code to both take the first guess given by the normalization template into account and at the same time allowing it to fine-tune the parameters towards optimal values for $\chi^2$ minimization. The output parameters are limited to the initial template parameters $\pm 350\,$K (for $T_\mathrm{eff}$), $\pm 0.2\,$dex ($\log g$) and $\pm 0.3\,$dex ($[M/H]$). We found these ranges through trial and error, and testing the method with multiple stars. They are a balance between the information given by the normalization template and the freedom required for the pipeline to find the best fit for a given spectrum.

\subsection{Error estimation \label{errors}}

Due to the way our pipeline is setup, we have two major possible error sources, one associated to a poor choice of normalization, and one connected to Turbospectrum and the derivation of the best fit synthetic spectrum for each star. In this section, we will try to quantify the errors associated to both of these sources across our M dwarf parameter space.

\subsubsection{Normalization template errors \label{NormErrors}}

As the shape of the characterized spectra, and consequent derived parameters, are strongly influenced by the normalization template used, we decided to test our pipeline by running multiple normalizations for the same observed star. A sample of 4 test stars was selected, covering a wide range of stellar parameters. For each test star, we ran the pipeline with all normalization templates with $\chi^2$ within a $10\%$ margin from the best templates found for each star, using the $\chi^2$ values as an indication that the template is appropriate for the studied star. We ran the pipeline 20 times for each normalization template, injecting Gaussian errors proportional to the $S/N$. The errors are injected so the deterministic pipeline does not return the same output parameters on every iteration, allowing us to know how the parameters can change across multiple observations of the same star. We do note here that these tests result in an underestimation of the actual errors, as errors affecting the continuum normalization will produce correlations across multiple pixels. Therefore these serve as a minimum estimate for the actual errors in our measurements. Table \ref{table:diffnormalizations} summarizes our results for a test sample of 4 stars, including the average and standard deviation obtained across the 20 runs of the pipeline for each normalization template per star.

The results displayed demonstrate that the output parameters can vary significantly with the template used for the normalization of the spectra. They also demonstrate the difficulty in assessing which template works best for a given star, as $\chi^2$ values can be very similar even for spectra with varying parameters. The standard deviation verified within the same template used also shows the consistency of the pipeline itself for a given spectrum, which is further explored in the following section.

\subsubsection{Errors associated with $S/N$}

In order to estimate the errors associated with the $S/N$ inherent to the spectra and the method used to find the best fit for each normalized spectra, a sample of 20 M dwarfs with different $S/N$ and stellar parameters was selected. The pipeline was ran 20 times per star, with Gaussian errors proportional to the reported $S/N$ of each star injected into the observed spectra before normalizing them with the method described in Section \ref{MdwarfNorm} Similar to the tests described in Section \ref{NormErrors}, these are also underestimations of the possible errors, as only uncertainties associated with individual pixels are taken into account. In order to have a more statistically relevant sample size, 80 additional iterations of the pipeline were run for each of 4 selected stars that cover a wide parameter range, for a total of 100 pipeline iterations per star. The results of these tests are summarized in Table \ref{table:synthmatchtest}.

As shown in the table, the output of the pipeline is very consistent across multiple iterations, with $T_\mathrm{eff}$ having standard deviations below 12\,K, $[M/H]<0.03$\,dex and $\log g<0.09\,$dex for all analyzed stars. There is no strong $S/N$ effect on the standard deviation measured, as the values remain very consistent across all stars in the sample. These tests suggest that most of the possible errors in the final results are a result of the templates used for the normalization of each observed spectrum, and showing that $\log g$ is the parameter with the largest associated error.

Given the errors measured in the displayed tests and the size of the grid used for the selection of the normalization template, we estimate the overall errors of our pipeline to be around $T_\mathrm{eff} \pm 100\,$K, $[M/H] \pm 0.1$\,dex and $\log g \pm 0.2\,$dex, with a large percentage of these errors coming from the inherent uncertainty in the choice of normalization template used for a given star. This choice is limited by the PARSEC evolutionary code, as it is the only one publicly available covering the M dwarf parameter space with a good range coverage. Another possible error source can be the line mask used for the $\chi^2$ calculation, as $\chi^2$ values can be very similar for different normalization templates and changes in the line mask can result in the selection of different templates as well. 

\begin{table*}
\caption{Average and standard deviation for stellar parameters measured for a sample of 4 stars, using different normalization templates and across 20 iterations of the pipeline for each star/normalization template combination. All displayed $T_\mathrm{eff}$ values are in K, while $\log g$ and [M/H] are in dex.}
\label{table:diffnormalizations}
\centering
\begin{tabular}{c c c c | c c c c}
\hline\hline 
{} & \multicolumn{3}{c}{Normalization} & \multicolumn{4}{c}{iSpec Output}\\

Star & $T_\mathrm{eff}$  & $\log g$ & $[M/H]$ &  $T_\mathrm{eff}$ &  $\log g$ & $[M/H]$ & $\chi^2$ \\ 
\hline
2M00243855+5119224	&	3700	&	4.9	&	-0.6	& $	3681	\pm	3	$ & $	5.16	\pm	0.02	$ & $	-0.49	\pm	0.01	$ & $	0.0455	\pm	0.0007	$ \\
2M00243855+5119224	&	3700	&	4.9	&	-0.8	& $	3671	\pm	2	$ & $	5.12	\pm	0.02	$ & $	-0.55	\pm	0.02	$ & $	0.0464	\pm	0.0006	$ \\
2M00243855+5119224	&	3700	&	5	&	-0.8	& $	3674	\pm	4	$ & $	5.14	\pm	0.02	$ & $	-0.54	\pm	0.02	$ & $	0.0462	\pm	0.0008	$ \\

\hline																							
2M01195227+8409327	&	3100	&	5	&	-0.2	& $	3113	\pm	7	$ & $	4.99	\pm	0.01	$ & $	-0.22	\pm	0.03	$ & $	0.196	\pm	0.002	$ \\
2M01195227+8409327	&	3100	&	5.1	&	-0.2	& $	3134	\pm	10	$ & $	5.07	\pm	0.01	$ & $	-0.29	\pm	0.04	$ & $	0.198	\pm	0.003	$ \\
2M01195227+8409327	&	3200	&	5	&	-0.2	& $	3201	\pm	4	$ & $	5.05	\pm	0.04	$ & $	-0.22	\pm	0.02	$ & $	0.201	\pm	0.003	$ \\
2M01195227+8409327	&	3200	&	5.1	&	-0.4	& $	3202	\pm	1	$ & $	5.13	\pm	0.01	$ & $	-0.38	\pm	0.01	$ & $	0.199	\pm	0.003	$ \\
\hline																							
2M03431519+5006558	&	3300	&	4.8	&	0	& $	3296	\pm	1	$ & $	4.93	\pm	0.01	$ & $	0.03	\pm	0.01	$ & $	0.191	\pm	0.002	$ \\
2M03431519+5006558	&	3300	&	4.9	&	0	& $	3301	\pm	1	$ & $	4.94	\pm	0.01	$ & $	0.03	\pm	0.01	$ & $	0.190	\pm	0.002	$ \\
2M03431519+5006558	&	3300	&	5	&	-0.2	& $	3275	\pm	10	$ & $	4.88	\pm	0.04	$ & $	-0.06	\pm	0.05	$ & $	0.204	\pm	0.004	$ \\
2M03431519+5006558	&	3400	&	4.9	&	-0.2	& $	3352	\pm	10	$ & $	5.02	\pm	0.03	$ & $	-0.05	\pm	0.03	$ & $	0.188	\pm	0.003	$ \\
\hline																							
2M23460112+7456172	&	3600	&	5.1	&	-0.8	& $	3578	\pm	4	$ & $	5.20	\pm	0.01	$ & $	-0.55	\pm	0.03	$ & $	0.069	\pm	0.001	$ \\
2M23460112+7456172	&	3700	&	5	&	-0.8	& $	3678	\pm	4	$ & $	5.20	\pm	0.01	$ & $	-0.64	\pm	0.03	$ & $	0.072	\pm	0.001	$ \\
2M23460112+7456172	&	3700	&	5.1	&	-1	& $	3675	\pm	4	$ & $	5.20	\pm	0.01	$ & $	-0.7	\pm	0.03	$ & $	0.072	\pm	0.001	$ \\
2M23460112+7456172	&	3800	&	4.9	&	-0.6	& $	3770	\pm	2	$ & $	5.10	\pm	0.01	$ & $	-0.68	\pm	0.02	$ & $	0.080	\pm	0.001	$ \\
2M23460112+7456172	&	3900	&	4.9	&	-0.8	& $	3845	\pm	3	$ & $	5.10	\pm	0.01	$ & $	-0.82	\pm	0.02	$ & $	0.088	\pm	0.001	$ \\
\hline
\end{tabular}
\end{table*}

\begin{table*}
\caption{Average and standard deviation for stellar parameters measured for multiple iterations for 20 selected stars representative of the full sample. 20 iterations were made for each of 16 test stars, while the code was made to run 100 iterations for 4 selected stars. All displayed $T_\mathrm{eff}$ values are in K, while $\log g$ and [M/H] are in dex. The ``Runs'' column displays the number of iterations made with the pipeline for each spectrum.}
\label{table:synthmatchtest}
\centering
\begin{tabular}{c c c c c c}
\hline\hline 
Star & $T_\mathrm{eff}$  & $\log g$ & $[M/H]$ & $S/N$ & Runs\\ 
\hline
2M00391896+5508132	& $	3329	\pm	5	$ & $ 	5.05	\pm	0.02	$ & $ 	-0.72	\pm	0.04	$ &	767	& 20\\
2M02465257+5619505	& $	3506	\pm	6	$ & $ 	4.69	\pm	0.02	$ & $ 	-0.29	\pm	0.02	$ &	346	& 20\\
2M04244284+4537062	& $	3859	\pm	4	$ & $ 	4.68	\pm	0.03	$ & $ 	-0.44	\pm	0.01	$ &	257	& 20\\
2M04422854+5818015	& $	3312	\pm	5	$ & $ 	5.05	\pm	0.01	$ & $ 	-0.26	\pm	0.03	$ &	615	& 20\\
2M05201152+2457212	& $	3667	\pm	7	$ & $ 	4.59	\pm	0.03	$ & $ 	-0.06	\pm	0.03	$ &	677	& 20\\
2M06070493+1403109	& $	3292	\pm	3	$ & $ 	5.08	\pm	0.02	$ & $ 	-0.21	\pm	0.01	$ &	258	& 20\\
2M06572462+0651440	& $	3307	\pm	8	$ & $ 	5.03	\pm	0.03	$ & $ 	-0.24	\pm	0.04	$ &	197	& 20\\
2M08050361+4121251	& $	3340	\pm	7	$ & $ 	4.91	\pm	0.03	$ & $ 	-0.68	\pm	0.06	$ &	222	& 20\\
2M10562960+4858264	& $	4063	\pm	12	$ & $ 	4.76	\pm	0.02	$ & $ 	0.21	\pm	0.01	$ &	328	& 20\\
2M11152550+0003159	& $	3424	\pm	4	$ & $ 	4.87	\pm	0.03	$ & $ 	-0.77	\pm	0.04	$ &	294	& 20\\
2M13552585+2556161	& $	4013	\pm	4	$ & $ 	4.51	\pm	0.01	$ & $ 	-0.48	\pm	0.01	$ &	177	& 20\\
2M14535251+1739448	& $	3801	\pm	3	$ & $ 	4.67	\pm	0.02	$ & $ 	-0.51	\pm	0.01	$ &	619	& 20\\
2M16495034+4745402	& $	3754	\pm	6	$ & $ 	4.69	\pm	0.02	$ & $ 	-0.01	\pm	0.01	$ &	1147	& 20\\
2M18055545+0316213	& $	3573	\pm	2	$ & $ 	4.63	\pm	0.01	$ & $ 	0	\pm	0.01	$ &	229	& 20\\
2M19004176+0310312	& $	3512	\pm	6	$ & $ 	4.95	\pm	0.03	$ & $ 	-0.77	\pm	0.09	$ &	170	& 20\\
2M23134861+1227072	& $	3560	\pm	2	$ & $ 	5	\pm	0.01	$ & $ 	0	\pm	0.01	$ &	172	& 20\\
\hline
2M03190939+0130543	& $	2972	\pm	3	$ & $ 	4.7	\pm	0.02	$ & $ 	0.22	\pm	0.01	$ &	161	 & 100 \\
2M04552111+5017249	& $	3882	\pm	3	$ & $ 	4.66	\pm	0.03	$ & $ 	-0.87	\pm	0.02	$ &	177	 & 100 \\
2M09301445+2630250	& $	3385	\pm	2	$ & $ 	4.67	\pm	0.02	$ & $ 	0.14	\pm	0.01	$ &	658	 & 100 \\
2M21400112+5408179	& $	3592	\pm	3	$ & $ 	5.03	\pm	0.02	$ & $ 	-0.55	\pm	0.02	$ &	563	 & 100 \\
\hline
\end{tabular}
\end{table*}

\begin{table*}
\caption{Output parameters for 6 stars with $T_\mathrm{eff} \leqslant  3000$\,K. All $T_\mathrm{eff}$ values are in Kelvin, $[M/H]$ in dex and $\log g$ in dex.}
\label{table:Coldest}
\centering
\begin{tabular}{c| c c c| c c c}
\hline\hline 
{} & \multicolumn{3}{c}{iSpec Output} & \multicolumn{3}{c}{Normalization}\\
Star & $T_\mathrm{eff}$ & $\log g$ & $\textrm{[M/H]}$ & $T_\mathrm{eff}$ & $\log g$ & $\textrm{[M/H]}$ \\
\hline
2M02081366+4949023	&	2886	&	4.80	&	0.25	&	2900	&	5.0	&	0.4	\\
2M05392474+4038437	&	2619	&	4.93	&	0.28	&	2600	&	5.1	&	0.4	\\
2M06481555+0326243	&	2831	&	5.17	&	-0.31	&	2800	&	5.2	&	-0.2	\\
2M13481341+2336486	&	3003	&	5.09	&	-0.21	&	3000	&	5.1	&	-0.2	\\
2M14562713+1755001	&	3117	&	5.05	&	-0.28	&	3100	&	5.0	&	-0.2	\\
2M20032651+2952000	&	3027	&	4.76	&	0.40	&	3000	&	4.9	&	0.4	\\
\hline
\end{tabular}
\end{table*}


\subsection{Summary}

This section contains a small summary of the steps required to derive M dwarf stellar parameters using our pipeline.

\begin{itemize}
\item 144 different synthetic spectra are generated using parameter combinations taken from PARSEC isochrones and expected for M dwarfs. We note that the MARCS models in this parameter space are provided in steps of 100K for $T_\mathrm{eff}$, $0.25\,dex$ for $[Fe/H]$, and 0.5\,dex for $\log g$, and we interpolate between them to create the grid whenever necessary. This interpolation is done entirely within \textit{iSpec}, and we have not edited that part of the code.
\item The observed combined spectrum for each sample star is downloaded from APOGEE, and normalized using each of the previously generated synthetic template spectra.
\item The best normalization template for each star is chosen based on a $\chi^2$ comparison between each template and its resulting normalized spectrum.
\item \textit{iSpec} \citep{blanco2014determining,ispec2019sbc}, as a shell for \textit{Turbospectrum} \citep{plez1998,plez2012turbospectrum}, is used to generate synthetic spectra from a set of starting values. These synthetic spectra are created using \textit{MARCS} \citep{MARCS} stellar atmospheric models and a custom line list.
\item Specific line masks are required for the synthesis of spectrum of stars with different spectral types, and they include the most relevant wavelength areas for spectral parameter determination.
\item The synthetic spectra are matched to the normalized observed ones through a $\chi^2$ minimization algorithm based on MPFIT \citep{markwardt2009non}. The algorithm runs on a list of previously selected wavelength regions defined as the line mask.
\item $\chi^2$ minimization is used to find the best match between a synthetic spectrum and a given observed one, interpolating the available \textit{MARCS} whenever necessary. Then, the spectral parameters used to generate the synthetic spectrum are taken as corresponding to the observed star.
\end{itemize}  

\section{Results \label{Results}}

\begin{figure}
\resizebox{\hsize}{!}{\includegraphics{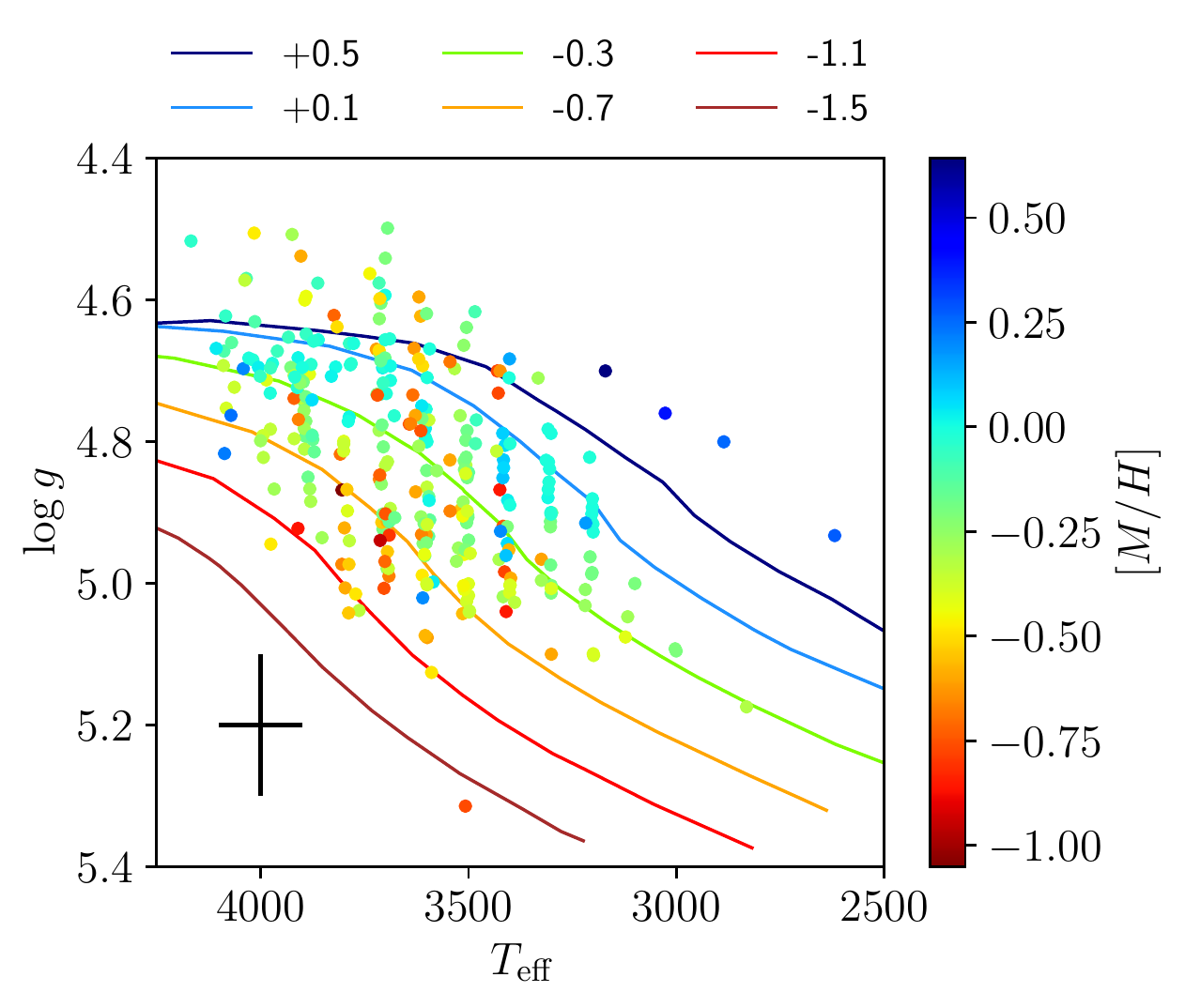}}
\caption{HR diagram with $T_\mathrm{eff}$ and $\log g$ values from iSpec output parameter distribution, with overplotted PARSEC isochrones \citep{bressan2012parsec} for an age of 5Gyr created with different $[M/H]$ values (scale is in dex). Points are color-coded based on the metallicity value derived for each star. Plotted parameters were obtained with APOGEE DR16}
\label{HR_Mdwarfs_Results}
\end{figure}

\begin{figure}
\resizebox{\hsize}{!}{\includegraphics{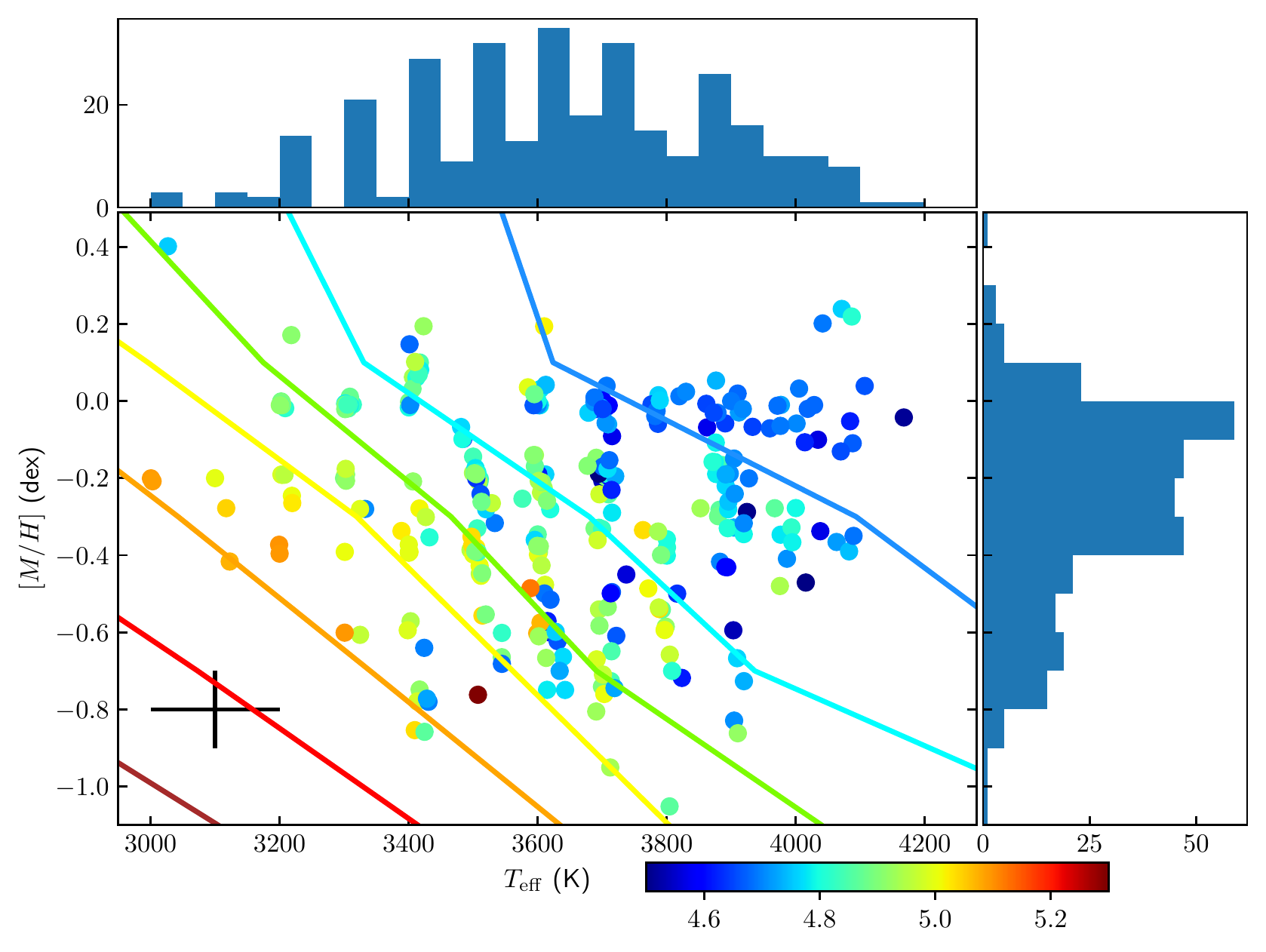}}
\caption{$T_\mathrm{eff}$ and $[M/H]$ from iSpec parameter distribution for stars in M dwarf sample, with overplotted PARSEC isochrones \citep{bressan2012parsec} for an age of 5Gyr created with different $\log g$ values (scale is in dex). Points are color-coded based on the surface gravity value derived for each star. Plotted parameters were obtained with APOGEE DR16}
\label{Teff_MH_ispec}
\end{figure}

\begin{figure}
\resizebox{\hsize}{!}{\includegraphics{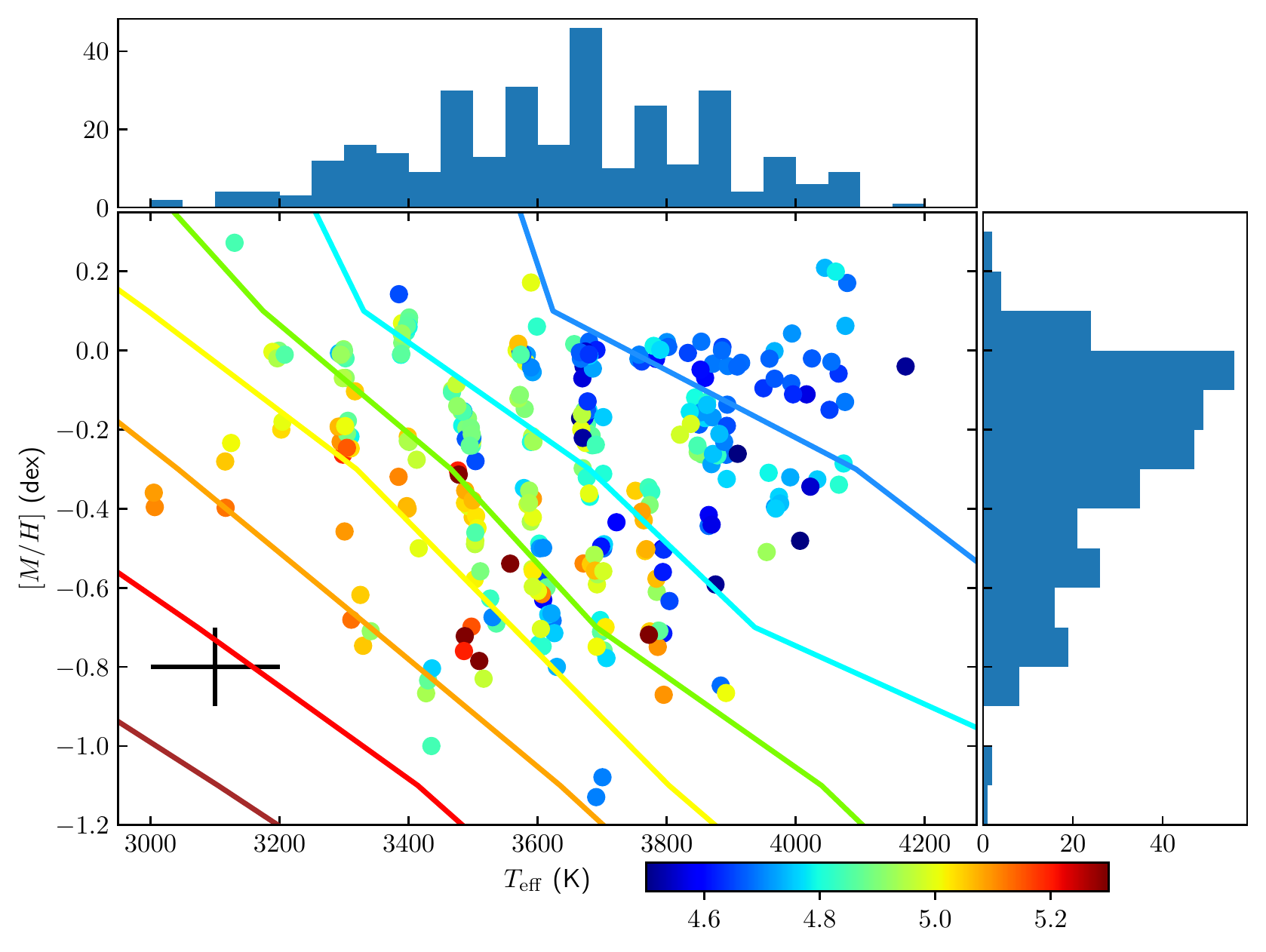}}
\caption{$T_\mathrm{eff}$ and $[M/H]$ from iSpec parameter distribution for stars in M dwarf sample, with overplotted PARSEC isochrones \citep{bressan2012parsec} for an age of 5Gyr created with different $\log g$ values (scale is in dex). Points are color-coded based on the surface gravity value derived for each star. Plotted parameters were obtained with APOGEE DR14}
\label{Teff_MH_ispecDR14}
\end{figure}

The spectra of the 313 M dwarfs in our sample were analyzed with the method detailed in Section \ref{method}, and the derived $T_\mathrm{eff}$, $\log g$ and $[M/H]$ are plotted in Figs. \ref{HR_Mdwarfs_Results} and \ref{Teff_MH_ispec}. The full results are available at the appendix, in Table \ref{table:MdwarfResultsDR16}. The errors included in the plots are $\pm 100\,$K for the $T_\mathrm{eff}$, $\log g \pm 0.2\,$dex, and $[M/H]\pm 0.1\,$dex. Fig.\ref{Teff_MH_ispecDR14} includes the parameters derived with APOGEE DR14, and the full list of parameters derived with that Data Release is included in Table \ref{table:MdwarfResultsDR14}, in the appendix.

The isochrone comparisons presented in Fig. \ref{HR_Mdwarfs_Results} show that, as expected, for most of the sample, our methodology's output $T_\mathrm{eff}$ and $\log g$ agree with the values predicted by the PARSEC models for M dwarfs. Despite this, the output $\log g$ values for a small number of stars is either overestimated or underestimated when compared to isochrone predictions. Some weak trends are also present across the results for the full sample in Fig. \ref{Teff_MH_ispec}, with thin columns of stars with very similar effective temperatures but different metallicity values. We do not know exactly what causes these trends. We find these values to cluster around multiples of 100\,K, and \textit{MARCS} models have spacings of 100\,K for $2500 \,$K$< T_\mathrm{eff} < 4000\,$K (compared with 250\,K for $4000 \,$K$< T_\mathrm{eff} < 8000\,$K), but further tests with the interpolation of the models have not shown any issues.

The lower limit of our analysis is around $T_\mathrm{eff} = 3000\,$K. From the sample of stars with ASPCAP $T_\mathrm{eff} < 3000\,$K, we found the normalization template with the lowest $\chi^2$ for 3 of them to converge to synthetic spectra with very high metallicity values ($[M/H]=0.4\,$dex) (see Table \ref{table:Coldest} for the full results derived for these stars). This correspond to the highest possible $[M/H]$ using our normalization method, and to the limit of the \textit{MARCS} models used to generate synthetic spectra. The output parameters for these three stars are very similar between them, and are outside the expected values for M dwarfs. We do not think these output parameters represent a reliable measurement for these stars, and they are included here only as a demonstration of the limitations of our method for stars with $T_\mathrm{eff} < 3000\,$K. We think that our pipeline's limitations are caused by our optimization of the line list to reproduce the spectra of stars with higher $T_\mathrm{eff}$, missing possible opacity sources for lower temperature stars, and the pipeline compensating these deficiencies by decreasing the $T_\mathrm{eff}$ and increasing the $[M/H]$ of the synthetic spectra. These effects increase the strength of the water lines present in the spectrum, but the resulting parameters will not be reliable.

We find the spectra of stars with $3000 <T_\mathrm{eff} < 3500\,$K to be especially challenging to synthetically reproduce due to the degeneracy in continuum absorption between multiple spectral parameters such as temperature, metallicity, and surface gravity. Absorptions in the continuum can be caused by any of these parameters, with the pipeline sometimes estimating lower metallicity and higher temperature values than the star has in reality.

\begin{landscape}

\begin{figure}
\resizebox{\hsize}{!}{\includegraphics{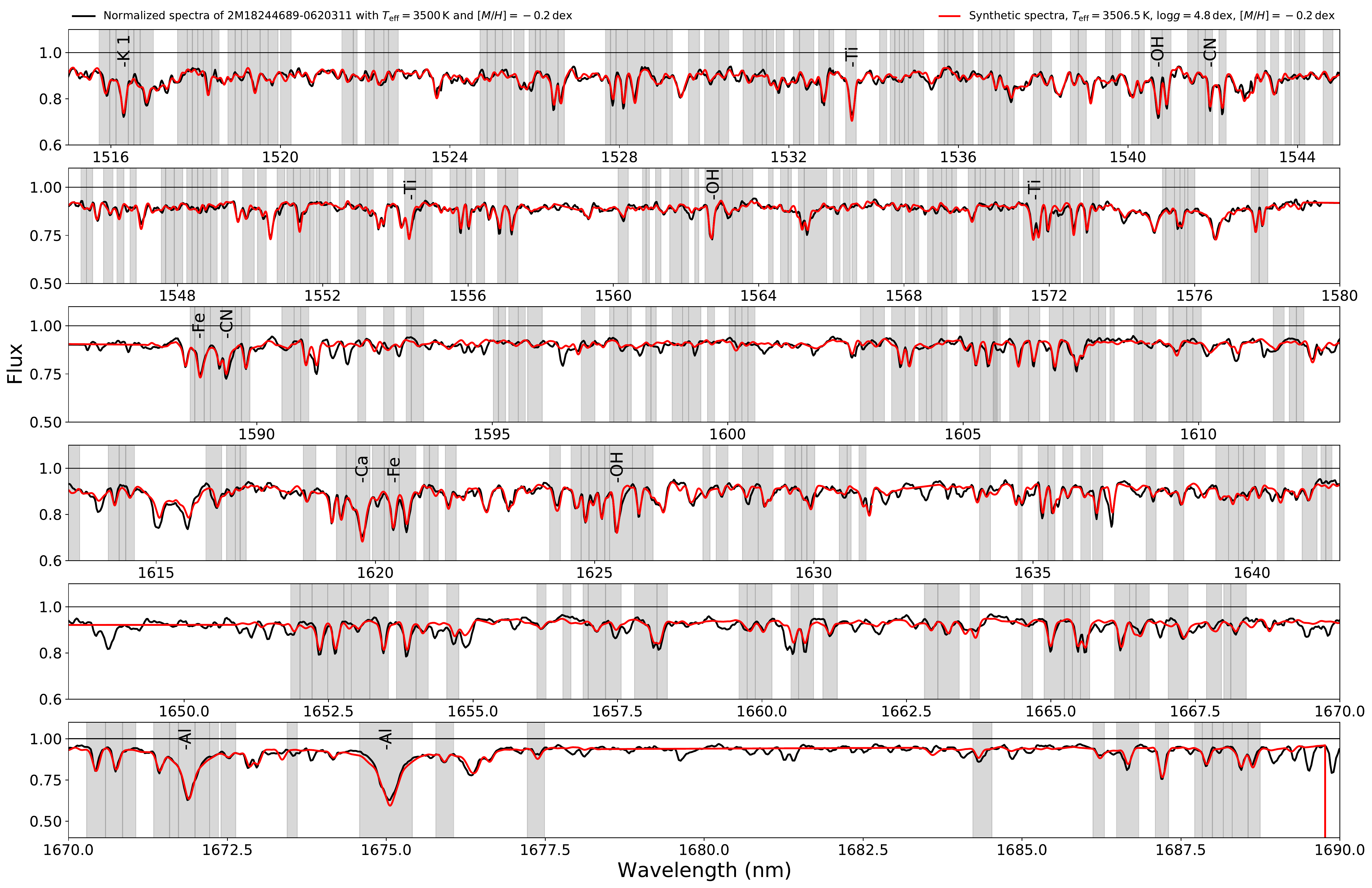}}
\caption{Comparison between APOGEE spectra of the star 2M18244689-0620311 (black, normalized using a synthetic spectra with $T_\mathrm{eff} = 3500\,$K, $\log g = 4.9$\,dex and $[M/H] = -0.2\,$dex) and the best fitting synthetic spectra (red, straight line) for APOGEE wavelength range. In gray highlight are the areas used for $\chi ^2 $ minimization by our pipeline's algorithm. The best fitting parameters derived were $T_\mathrm{eff}=3507\pm100$\,K,$\log g=4.80\pm0.2$\,dex, $[M/H] = -0.19\pm0.1$\,dex. The available ASPCAP parameters are $T_\mathrm{eff}=3600\pm59$\,K, $\log g=5.09\pm0.12$\,dex and $[M/H] = -0.06\pm0.01$\,dex. This star was characterized in \cite{Souto2020} as having $T_\mathrm{eff}=3376$\,K,$\log g=4.77$\,dex, $[M/H] = +0.21$\,dex. }
\label{Mdwarf_3}
\end{figure}

\end{landscape}

We present an example synthesized stellar spectra for a star with $T_\mathrm{eff} \sim 3500\,$K in Fig. \ref{Mdwarf_3}. Here, the spectra of star 2M18244689-0620311 (BD-06 4756B \footnote{\url{http://simbad.u-strasbg.fr/simbad/sim-id?Ident=2Mass+j18244689-0620311}}) is displayed, showing both the observed (black) and best matching spectra (red). This star was characterized in \cite{Souto2020} as having $T_\mathrm{eff}=3376$\,K,$\log g=4.77$\,dex, $[M/H] = +0.21$\,dex. The available ASPCAP parameters for this star are $T_\mathrm{eff}=3502$\,K, $\log g=4.63$\,dex, and $[M/H] = -0.17$\,dex. The displayed spectrum for the star has been obtained by normalizing the APOGEE observed one with a synthetic spectrum with $T_\mathrm{eff} = 3500\,$K, $\log g = 4.9$\,dex and $[M/H] = -0.2\,$dex, and the derived spectroscopical parameters for this star are $T_\mathrm{eff}=3484\pm100$\,K,$\log g=4.85\pm0.2$\,dex, $[M/H] = -0.15\pm0.1$\,dex.

The presence of a wide water line band is noticeable across the full spectrum, especially the full first order (two top rows), resulting in a jagged and depressed continuum down to around 0.9 instead of the expected 1.0. The modeling of these water lines is fundamental for the creation of an accurate synthetic spectra of late M dwarfs ($T_\mathrm{eff}<3500$\,K). This star is slightly below solar metallicity, still having pronounced molecular and elemental lines. The effect can be noticed in the strong Al line around 1675\,nm, a line accurately matched by the synthetic spectrum. The synthesized spectrum matches the observed one across the full wavelength range, with lines of multiple elements and molecules being correctly synthesized. This fact, combined with the relative agreement of our pipeline's derived parameter with available literature analysis, increases our confidence in the derived parameters for this star.

Additional examples of output spectra obtained by our pipeline are displayed in the appendix, in Figs. \ref{Mdwarf_1} to \ref{Mdwarf_6}, with more stars across our parameter space.

\section{Analysis \label{Discussion}}

\subsection{ASPCAP comparison \label{ASPCAP_comp}}

\begin{figure*}
\resizebox{\hsize}{!}{\includegraphics[height = 19cm,keepaspectratio=true]{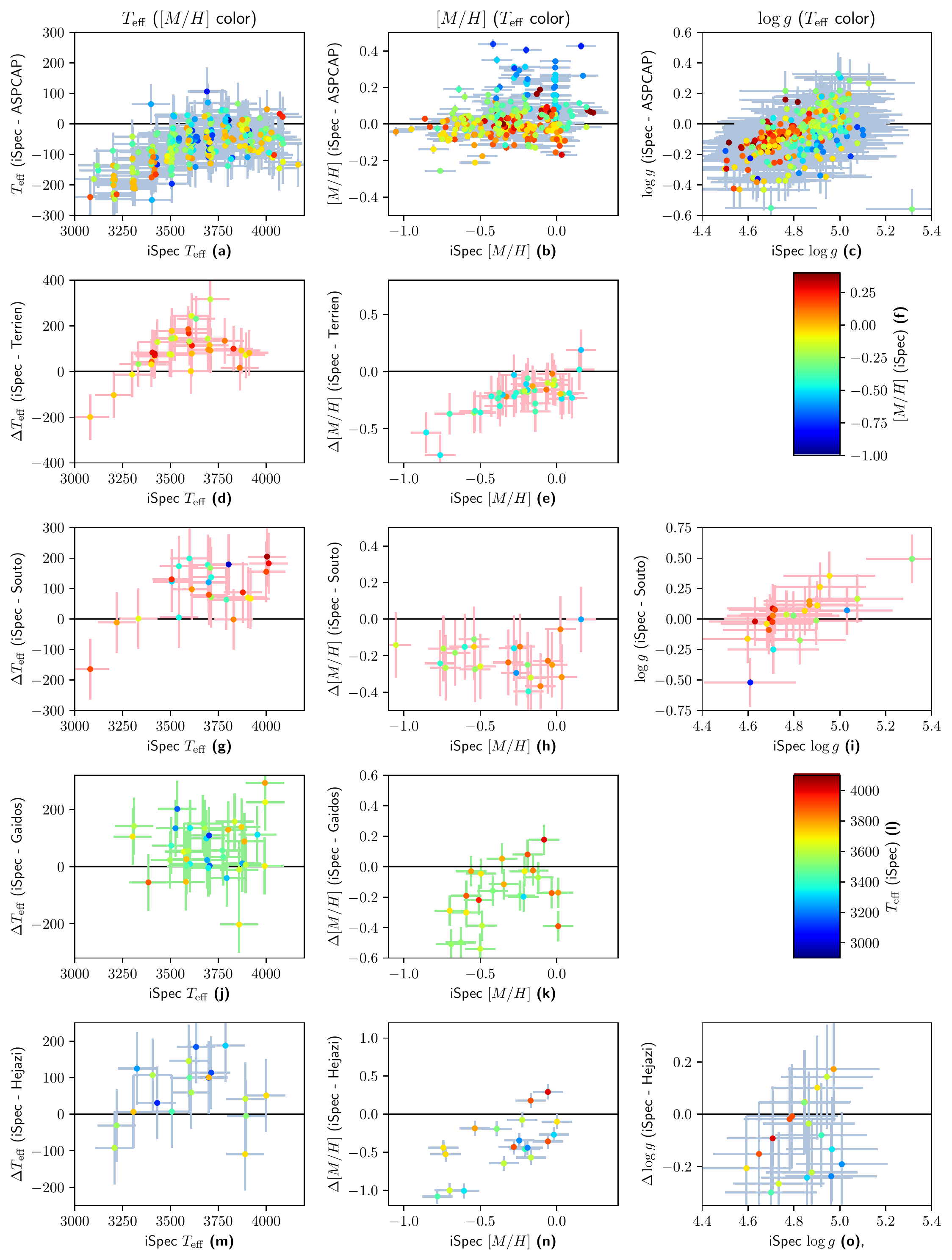}}
\caption{Comparisons between our output parameters and ones available in literature, colored accordingly to either our method's measured $[M/H]$ ($T_\mathrm{eff}$ plots, left column, see scale at (f)) or $T_\mathrm{eff}$ for each star ($[M/H]$, center column, and $\log g$, right column, see color scale at (l)). From the top to bottom: comparisons with ASPCAP (a, b, c), \cite{terrien2015near} (d, e), \cite{Souto2020} (g, h, i), \cite{gaidos2014trumpeting} (j, k), \cite{Hejazi2019Chemical} (m, n, o).  Temperature values are displayed in Kelvin, while metallicity and surface gravity are in dex.}
\label{All_comparisons}
\end{figure*}
\twocolumn

\begin{table}
\caption{Median differences between our results and literature parameters. All displayed ASPCAP parameters ($T_\mathrm{eff}$, $[M/H]$ and $\log g$) are from their calibrated values. Median difference is calculated by subtracting literature parameters from our results. SD stands for standard deviation and MAD for Median Absolute Deviation. Temperature values are displayed in Kelvin, while metallicity and surface gravity are in dex.}
\label{table:All_comp_table}
\centering
\begin{tabular}{c c c c c}
\hline\hline 
Parameter & Stars & Median & SD & MAD \\ 
\hline
\multicolumn{5}{c}{ASPCAP} \\
\hline
$T_\mathrm{eff}$ & 283 & -63 & 70 & 58 \\
$[M/H]$ & 283 & +0.0 & 0.15 & 0.08 \\
$\log g$ & 283 & -0.07 & 0.14 & 0.11 \\
\hline
\multicolumn{5}{c}{\cite{terrien2015near}} \\
\hline
$T_\mathrm{eff}$ & 45 & +92 & 86 & 57 \\
$[M/H]$ & 45 & -0.19 & 0.15 & 0.08 \\
\hline
\multicolumn{5}{c}{\cite{Souto2020}} \\
\hline
$T_\mathrm{eff}$ & 24 & +109 & 85 & 77 \\
$[M/H]$ & 24 & -0.23 & 0.09 & 0.11 \\
$\log g$ & 24 & +0.03 & 0.20 & 0.11 \\
\hline
\multicolumn{5}{c}{\cite{gaidos2014trumpeting}} \\
\hline
$T_\mathrm{eff}$ & 40 & +56 & 91 & 81 \\
$[M/H]$ & 24 & -0.17 & 0.19 & 0.19\\
\hline
\multicolumn{5}{c}{\cite{Hejazi2019Chemical}} \\
\hline
$T_\mathrm{eff}$ & 19 & +59 & 81 & 78 \\
$[M/H]$ & 19 & -0.43 & 0.36 & 0.32 \\
$\log g$ & 19 & -0.09 & 0.14 & 0.19 \\
\hline
\end{tabular}
\end{table}

As mentioned in Section \ref{SampleParams}, there are no available final ASPCAP spectroscopic parameters for all stars in our sample. From the full 313 star sample, we have ASPCAP calibrated values for $T_\mathrm{eff}$, $[M/H]$, and $\log g$ for 283 of them. Therefore, all comparisons shown and discussed here will focus on those 283 stars. We do not show raw/spectroscopic parameters for consistency reasons. Fig.\ref{All_comparisons} (a) shows a comparison between the $T_\mathrm{eff}$ values for our sample M dwarfs as derived by our pipeline and the calibrated values published by ASPCAP for the same stars, (b) presents the same comparison for ASPCAP calibrated $[M/H]$ values, and (c) shows a comparison between the $\log g$ values derived with our pipeline and the calibrated ones published by ASPCAP. Errors included are estimated based on our normalization grids and are $\pm 100\,$K for the $T_\mathrm{eff}$, $\pm 0.1\,$dex for $[M/H]$ and $\pm 0.2\,$dex for $\log g$ (see section \ref{errors}). The points are colored accordingly to other parameters measured by our method in order to highlight any correlation or trend between errors across multiple output parameters.

In Fig. \ref{All_comparisons} (a), and relative to $\Delta T_\mathrm{eff}$, we find a small trend of more negative $\Delta T_\mathrm{eff}$ for stars with $T_\mathrm{eff} < 3500\,$K. For stars above this $T_\mathrm{eff}$, we find very similar values between our analysis and the parameters published by ASPCAP. Additionally, we find no correlation between $\Delta T_\mathrm{eff}$ and each star's metallicity values. Overall, we find this parameter to have a good agreement with ASPCAP published results, with the median difference, SD and MAD all being within our error margins. 

As for Fig. \ref{All_comparisons} (b) and $[M/H]$, a trend is present here, as we find a negative $\Delta [M/H]$ for the stars in our sample with $T_\mathrm{eff}>3400\,$K and a positive one for stars with $T_\mathrm{eff}<3400\,$K. This translates to an overestimation of metallicity for the colder stars in our sample, when compared to ASPCAP values. Both lower $T_\mathrm{eff}$ and increasing $[M/H]$ can have a similar effect on the spectral shape, lowering the continuum position due to an increased strength of the water molecular lines. Therefore, metallicity overestimations can be caused by normalization errors and/or correspond to effective temperature underestimations. We have to note that we are using the $T_\mathrm{eff}$ values determined by our pipeline to color the points.

For $\log g$, Fig. \ref{All_comparisons} (c) shows that a small trend is present in the data, with negative $\Delta \log g$ values for stars with lower $\log g$. The overall distribution is not very different, with median differences of +0.07\,dex between our parameters and the ones published by ASPCAP, although with SD and MAD around our uncertainty levels. We find a wide range of differences in $\log g$, with the comparisons varying between -0.6\,dex and +0.4\,dex. Comparisons with the isochrones in Fig. \ref{HR_Mdwarfs_Results} show that our values for this parameter are within the expectations for main-sequence stars around these temperatures. A direct comparison between Figs. \ref{Mdwarfs_ASPCAP_params}, \ref{Teff_MH_ispec} and \ref{Teff_MH_ispecDR14} shows how the ASPCAP calibrated $\log g$ values differ from our derived parameters for our M dwarf sample. While our values using data from DR16 follow the isochrones approximately, both the results with DR14 and their parameters seem to vary wildly, especially for M dwarfs with $T_\mathrm{eff}<3800\,$K / $\log g > 4.9\,$dex. This does seem to indicate that our method provides more plausible surface gravity values, given our current stellar modeling knowledge reflected in the overplotted PARSEC isochrones. It also shows how the spectra provided by APOGEE improved between DR14 and DR16, lending credence to the more recent Data Release as being of higher quality than the previous one.

Overall, we find that the best agreement between our parameters and the ones published by ASPCAP is found for $[M/H]$, despite some trends being present alongside $T_\mathrm{eff}$ and $\log g$. All of these comparisons are made with calibrated ASPCAP parameters. These should not be ignored, as they can show inherent bias in the method and the output parameters.

\subsection{Other literature comparisons}

\subsubsection{Comparison with \cite{terrien2015near}}

Fig. \ref{All_comparisons} (d) and (e) show a comparison between the parameters derived with our pipeline and the ones published in \cite{terrien2015near} for 45 stars in common between both samples. Table \ref{table:All_comp_table} shows the median difference, standard deviation and mean absolute deviation found when comparing both distributions. As \cite{terrien2015near} does not publish values for $\log g$, we kept the space reserved for the plot for that parameter (right column) empty. We find the differences between them to be above the uncertainty level, with the SD and MAD of the $[M/H]$ values being slightly above (0.15 and 0.08\,dex vs 0.10\,dex) and $T_\mathrm{eff}$ slightly below (86\,K and 57\,K vs 100\,K).

In the case of $T_\mathrm{eff}$, we find no large-scale trend or deviation between both parameter distributions, with $\Delta T_\mathrm{eff} < 250\,$K for all compared stars and with no correlation with $[M/H]$. However, the fact that we find $\Delta T_\mathrm{eff}\geq 0\,$K for almost the full sample, as well as a median difference of $+92\,$K, points towards small biases in either parameter distribution. As for $\Delta [M/H]$, it has a wider distribution, going up to $\pm 0.3\,$dex, and a median difference of $-0.19\,$dex can point towards systematic trends in our output data. There is also a linear trend towards negative $\Delta [M/H]$ for stars with $[M/H] < -0.2$\,dex and $T_\mathrm{eff} \sim 3300\,$K.

\subsubsection{Comparison with Souto's parameters}

Figs. \ref{All_comparisons} (g), (h), and (i) show a comparison between the parameters derived by our pipeline and the ones published in \cite{souto2017, souto2018stellar, Souto2020} for 24 M dwarfs. We should note that, as Souto et al. published $[Fe/H]$ for their characterized stars and our method measures $[M/H]$, that discrepancy can explain some of the differences. The differences between results published by both methods are also summarized in Table \ref{table:All_comp_table}. 

We find clear differences between both parameter distributions, with lower metallicity values ($\Delta [M/H] = -0.23\,$dex) and higher temperatures ($\Delta T_\mathrm{eff} = +109\,$K) for our stars. These trends fall in line with the results for the previously compared \cite{terrien2015near} results, with negative $\Delta [M/H]$ and positive $\Delta T_\mathrm{eff}$, with the metallicity differences being above our uncertainty levels. Like the previous comparison, we find no cross-parameter trends, as the $[M/H]$ differences seem to be independent of $T_\mathrm{eff}$ and vice versa.

We also find a weak linear trend with $\log g$ itself, as we find lower $\Delta \log g$ for stars with $\log g < 4.8\,$dex, and a higher $\Delta \log g$ for stars with $\log g>4.8\,$dex (with a correlation coefficient of $\rho = 0.71$).

\subsubsection{Comparison with \cite{gaidos2014trumpeting}}

We present, in Fig. \ref{All_comparisons} (j), a comparison between our $T_\mathrm{eff}$ results and the ones published in \cite{gaidos2014trumpeting} for 40 stars in common between both samples. Fig. \ref{All_comparisons} (k) shows a similar comparison for $[M/H]$ for the 24 stars with that parameter in the common sample. Similar to the case of \cite{terrien2015near}, we have no reference value for $\log g$ from this source, so the space for the plot for that parameter (right column) is kept empty. Additionally, Table \ref{table:All_comp_table} presents the median, standard deviation and mean absolute deviation found when comparing both parameter distributions.

Similar to the previous comparisons, we find trends towards both positive $\Delta T_\mathrm{eff}$ and negative $\Delta [M/H]$. We find the standard (91\,K) and mean absolute deviations (81\,K) to be slightly around our estimated uncertainties for temperature, while the median differences (+56\,K) are below them. As for metallicity, they are slightly above our uncertainty levels. There is a small trend towards lower $\Delta [M/H]$ for colder objects, but since the number of stars is so small it might be just a statistical artifact.

\subsubsection{Comparison with \cite{Hejazi2019Chemical}}

We present, in Figs. \ref{All_comparisons} (m), (n) and (o), a comparison between our output values and the ones published in \cite{Hejazi2019Chemical} for 19 stars in common between both samples. Additionally, Table \ref{table:All_comp_table} presents the differences between results obtained by both methods. We find positive $\Delta T_\mathrm{eff}$, with standard (81\,K) and mean absolute (78\,K) deviations around our estimated uncertainties for the parameter. We also find strong trends in both $\log g$ and $[M/H]$ distributions, with deviations significantly above our estimated uncertainty values.

As for discrepancies in $T_\mathrm{eff}$, Fig. \ref{All_comparisons} (m) shows that the $\Delta T_\mathrm{eff}$ distribution between both methods is not very large, with most stars having a $\Delta T_\mathrm{eff} \sim \pm 120\,$K. The strongest outlier is star 2M07404603+3758253, which we find to have $T_\mathrm{eff} = 3325\,$K and $[M/H] = -0.62\,$dex, while \cite{Hejazi2019Chemical} published $T_\mathrm{eff} = 3200\,$K and $[M/H] = +0.4\,$dex for the same star \footnote{A plot comparing our best synthetic fit to the observed APOGEE spectra is included in the Appendix as Fig. \ref{Mdwarf_7}.}. 

A $[M/H]$ comparison shows a linear trend in Fig. \ref{All_comparisons} (n), with $\Delta [M/H] \sim -1.0$\,dex for metal-poor stars, and $\Delta [M/H] \sim 0.0$\,dex for stars around solar metallicity. We also find significantly lower values for metallicity than \cite{Hejazi2019Chemical}, with an average $\Delta [M/H] = -0.43\,$dex. Despite their reported uncertainties of $\pm0.22\,$dex, this is a significant difference. Similar to the comparison with \cite{gaidos2014trumpeting} (Fig.\ref{All_comparisons} (k)) and unlike the ASPCAP comparison (Fig.\ref{All_comparisons} (b), there is a trend towards lower $\Delta [M/H]$ for colder objects. Some possible explanations can be the lower resolution of their analysis and the fact that it was done in the optical and not in the near-infrared, but the fact that the trends are very similar across all literature comparisons presented here lead us to believe in the presence of an inherent bias in our pipeline towards low $[M/H]$ output parameters.

For $\log g$, we find a linear trend towards measuring lower $\Delta \log g$ for stars with $\log g < 4.9\,$dex, and a higher $\Delta \log g$ for stars with $\log g>4.9\,$dex. This is very similar to the $\Delta \log g$ plot for our Souto values comparison (see Fig.\ref{All_comparisons} (i)) and can point towards some issues with the $\log g$ determined by our method.

\subsubsection{Comparison with \cite{rajpurohit2018apogee}}



\begin{figure*}
\resizebox{\hsize}{!}{\includegraphics{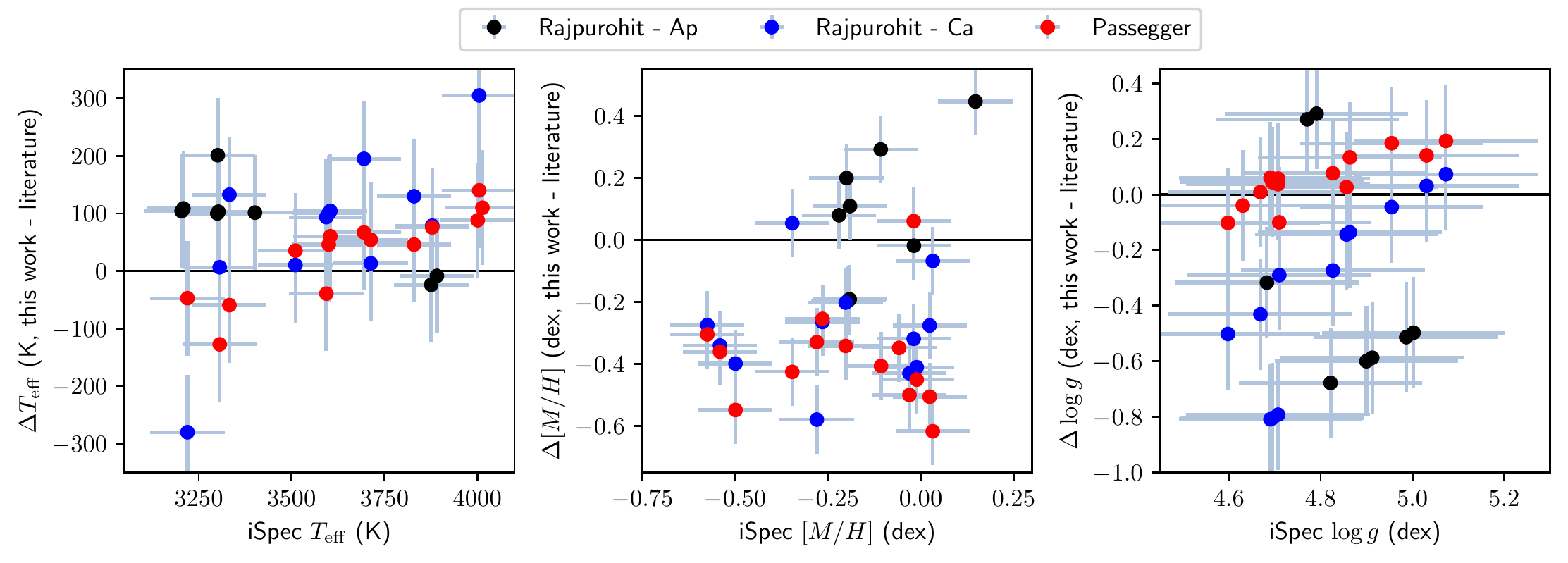}}
\caption{Comparison in derived parameters between our method and the eight stars in common with \cite{rajpurohit2018apogee} (APOGEE) and the twelve stars in common with \cite{rajpurohit2018carmenes} (CARMENES), and fourteen with \cite{passegger2019carmenes} (our pipeline - literature).}
\label{Raj_Pass_comp}
\end{figure*}



We display a comparison between our output parameters and the results published by \cite{rajpurohit2018apogee} for 8 stars in common in Fig. \ref{Raj_Pass_comp}. Due to the low number of stars compared, no real statistical analysis can be made from this comparison. We find some small differences in the results across the stellar sample, with the clearest example being star 2M06320207+3431132, with an output $T_\mathrm{eff}=3465\,$K (our results) vs 3200\,K. Similar to the previous comparisons, we find an overall positive, but close to our uncertainty levels, $\Delta T_\mathrm{eff}$. Unlike the previous results, we actually find a positive $\Delta [M/H]$ as well, while the largest difference between our results and theirs is in surface gravity, with some stars having $\Delta \log g\sim-0.6\,$dex.

\subsubsection{Comparison with \cite{rajpurohit2018carmenes} and \cite{passegger2019carmenes}}

As our sample and both \cite{rajpurohit2018carmenes} and \cite{passegger2019carmenes}'s have only 12 stars and 14 stars in common, respectively, we compared our results with the parameters for each of them in Fig \ref{Raj_Pass_comp}. No overall trends are present in $ T_\mathrm{eff}$, with the exception of a single star, 2M11474440+0048164, that has \cite{rajpurohit2018carmenes} $T_\mathrm{eff}=3500\,$K and our analysis output is $T_\mathrm{eff}=3202\,$K. This star, however is characterized in \cite{passegger2019carmenes} as having $T_\mathrm{eff}=3267\,$K, and by \cite{souto2018stellar} as having $T_\mathrm{eff}=3231\,$K. Both these literature values are much closer to our results, explaining the results of \cite{rajpurohit2018carmenes} as an outlier. As for metallicity, we find $\Delta [M/H] \sim - 0.3\,$dex, in line with other literature comparisons, and keep in mind that the comparison with \cite{passegger2019carmenes} is based on their values for $[Fe/H]$ rather than $[M/H]$. As for $\log g$, we find our output parameters to be very similar to the ones published by \cite{passegger2019carmenes}, but some differences are present in the comparison with \cite{rajpurohit2018carmenes}, with our values being up to $0.8\,dex$ below theirs for 3 stars in common. However, similarly to the $ T_\mathrm{eff}$ for star 2M11474440+0048164, the $\log g$ values published by \cite{passegger2019carmenes} for these stars are also much closer to our measurements, with a $\Delta \log g \sim 0.05\,$dex.

\section{Conclusions \label{Conclusions}}

Building on the method previously presented in \cite{sarmento2020derivation}, we derived stellar atmospheric parameters for 313 M dwarfs with $3000\,$K$ < T_\mathrm{eff}\pm100\,$K$ < 4200\,$K, $4.5\,$dex$< \log g\pm0.2\,$dex$ < 5.3\,$dex and $-1.05\,$dex $ < [M/H]\pm0.1\,$dex$ < 0.56\,$dex. The pipeline uses \textit{iSpec} as a spectroscopic framework to control \textit{Turbospectrum} code, and requires both a complete line list at the studied wavelength and a line mask tailored for the spectral type of the analyzed stars. We include both of these resources with the paper, as well as all the derived stellar parameters, for future reference.

A series of literature comparisons demonstrates the difficulty of finding accurate parameters for M dwarfs. Positive $\Delta T_\mathrm{eff}$ values are found across multiple literature comparisons, except for the ASPCAP values, although usually within uncertainty levels. However, the additional presence of negative $\Delta [M/H]$ values in most of our literature comparisons suggest the existence of issues with the temperature and metallicity determination in our pipeline, as the effects of changes in these parameters in the M dwarf regime are not independent. We also note that our results are also dependent on the assumptions of the PARSEC evolutionary code. More analysis needs to be done on this subject.

Despite the existence of these trends, the strong matching between synthesized and observed spectra for a wide range of M dwarf stars shows the power of the pipeline as a method for parameter determination. Future works could apply the method to Near-Infrared spectra observed with other instruments with better resolution than APOGEE, such as CARMENES \citep[$R=80\,000-100\,000$,][]{quirrenbach2014carmenes}, GIANO \citep[$R\sim50\,000$,][]{origlia2014high}, SPIROU \citep[$R\sim75\,000$,][]{artigau2014spirou}, NIRPS \citep[$R\sim90\,000-100\,000$][]{wildi2017nirps}, and CRIRES+ \citep[$R\sim100\,000$,][]{dorn2014crires+}. Expanding the characterized parameter space, through an analysis of a larger M dwarf sample, together with an expansion to more complete and detailed molecular line lists, are other possibilities to improve the performance of our method.

\begin{acknowledgements}
 This work was supported by FCT - Fundação para a Ciência e a Tecnologia through national funds (PTDC/FIS-AST/28953/2017, 
 PTDC/FIS-AST/7073/2014, PTDC/FIS-AST/32113/2017, UID/FIS/04434/2013) 
 and by FEDER - Fundo Europeu de Desenvolvimento Regional through COMPETE2020 - 
 Programa Operacional Competitividade e Internacionalização 
 (POCI-01-0145-FEDER-028953, POCI-01-0145-FEDER-016880, POCI-01-0145-FEDER-032113, POCI-01-0145-FEDER-007672).
This work was supported by Fundação para a Ciência e a Tecnologia (FCT) through the research grants UID/FIS/04434/2019, UIDB/04434/2020 and UIDP/04434/2020.
 This research has made use of NASA’s Astrophysics Data System.
 P.S. acknowledges the support by the Bolsa de Investigação PD/BD/128050/2016.
 E.D.M. acknowledges the support by the Investigador FCT contract IF/00849/2015/CP1273/CT0003 and in the form of an exploratory project with the same reference.
 B.R-A acknowledges funding support from FONDECYT through grant 11181295.
\end{acknowledgements}

\bibliographystyle{aa}
\bibliography{notes2.bib}

\begin{thebibliography}{64}
\expandafter\ifx\csname natexlab\endcsname\relax\def\natexlab#1{#1}\fi

\bibitem[{{Ahumada} {et~al.}(2019){Ahumada}, {Allende Prieto}, {Almeida},
  {Anders}, {Anderson}, {Andrews}, {Anguiano}, {Arcodia}, {Armengaud},
  {Aubert}, {Avila}, {Avila-Reese}, {Badenes}, {Balland}, {Barger},
  {Barrera-Ballesteros}, {Basu}, {Bautista}, {Beaton}, {Beers}, {Benavides},
  {Bender}, {Bernardi}, {Bershady}, {Beutler}, {Moni Bidin}, {Bird}, {Bizyaev},
  {Blanc}, {Blanton}, {Boquien}, {Borissova}, {Bovy}, {Brandt}, {Brinkmann},
  {Brownstein}, {Bundy}, {Bureau}, {Burgasser}, {Burtin}, {Cano-Diaz},
  {Capasso}, {Cappellari}, {Carrera}, {Chabanier}, {Chaplin}, {Chapman},
  {Cherinka}, {Chiappini}, {Choi}, {Chojnowski}, {Chung}, {Clerc}, {Coffey},
  {Comerford}, {Comparat}, {da Costa}, {Cousinou}, {Covey}, {Crane}, {Cunha},
  {da Silva Ilha}, {Dai}, {Damsted}, {Darling}, {Horta Darrington}, {Davidson},
  {Davies}, {Dawson}, {De}, {de la Macorra}, {De Lee}, {Queiroz}, {Deconto
  Machado}, {de la Torre}, {Dell'Agli}, {du Mas des Bourboux},
  {Diamond-Stanic}, {Dillon}, {Donor}, {Drory}, {Duckworth}, {Dwelly},
  {Ebelke}, {Eftekharzadeh}, {Davis Eigenbrot}, {Elsworth}, {Eracleous},
  {Erfanianfar}, {Escoffier}, {Fan}, {Farr}, {Fernandez-Trincado}, {Feuillet},
  {Finoguenov}, {Fofie}, {Fraser-McKelvie}, {Frinchaboy}, {Fromenteau}, {Fu},
  {Galbany}, {Garcia}, {Garcia-Hernandez}, {Garma Oehmichen}, {Ge}, {Geimba
  Maia}, {Geisler}, {Gelfand }, {Goddy}, {Le Goff}, {Gonzalez-Perez},
  {Grabowski}, {Green}, {Grier}, {Guo}, {Guy}, {Harding}, {Hasselquist},
  {Hawken}, {Hayes}, {Hearty}, {Hekker}, {Hogg}, {Holtzman}, {Hou}, {Hsieh},
  {Huber}, {Hunt}, {Ider Chitham}, {Imig}, {Jaber}, {Jimenez Angel}, {Johnson},
  {Jones}, {Jonsson}, {Jullo}, {Kim}, {Kinemuchi}, {Kirkpatrick}, {Kite},
  {Klaene}, {Kneib}, {Kollmeier}, {Kong}, {Kounkel}, {Krishnarao}, {Lacerna},
  {Lan}, {Lane}, {Law}, {Leung}, {Lewis}, {Li}, {Lian}, {Lin}, {Long},
  {Longa-Pena}, {Lundgren}, {Lyke}, {Mackereth}, {MacLeod}, {Majewski},
  {Manchado}, {Maraston}, {Martini}, {Masseron}, {Masters}, {Mathur},
  {McDermid}, {Merloni}, {Merrifield}, {Meszaros}, {Miglio}, {Minniti},
  {Minsley}, {Miyaji}, {Gohar Mohammad}, {Mosser}, {Mueller}, {Muna},
  {Munoz-Gutierrez}, {Myers}, {Nadathur}, {Nair}, {Correa do Nascimento},
  {Nevin}, {Newman}, {Nidever}, {Nitschelm}, {Noterdaeme}, {O'Connell},
  {Olmstead}, {Oravetz}, {Oravetz}, {Osorio}, {Pace}, {Padilla},
  {Palanque-Delabrouille}, {Palicio}, {Pan}, {Pan}, {Parker}, {Paviot},
  {Peirani}, {Pena Ramrez}, {Penny}, {Percival}, {Perez-Fournon},
  {Perez-Rafols}, {Petitjean}, {Pieri}, {Pinsonneault}, {Poovelil}, {Povick},
  {Prakash}, {Price-Whelan}, {Raddick}, {Raichoor}, {Ray}, {Barboza Rembold},
  {Rezaie}, {Riffel}, {Riffel}, {Rix}, {Robin}, {Roman-Lopes}, {Roman-Zuniga},
  {Rose}, {Ross}, {Rossi}, {Rowlands}, {Rubin}, {Salvato}, {Sanchez},
  {Sanchez-Menguiano}, {Sanchez-Gallego}, {Sayres}, {Schaefer}, {Schiavon},
  {Schimoia}, {Schlafly}, {Schlegel}, {Schneider}, {Schultheis}, {Schwope},
  {Seo}, {Serenelli}, {Shafieloo}, {Shamsi}, {Shao}, {Shen}, {Shetrone},
  {Shirley}, {Silva Aguirre}, {Simon}, {Skrutskie}, {Slosar}, {Smethurst},
  {Sobeck}, {Cervantes Sodi}, {Souto}, {Stark}, {Stassun}, {Steinmetz},
  {Stello}, {Stermer}, {Storchi-Bergmann}, {Streblyanska}, {Stringfellow},
  {Stutz}, {Suarez}, {Sun}, {Taghizadeh-Popp}, {Talbot}, {Tayar}, {Thakar},
  {Theriault}, {Thomas}, {Thomas}, {Tinker}, {Tojeiro}, {Hernandez Toledo},
  {Tremonti}, {Troup}, {Tuttle}, {Unda-Sanzana}, {Valentini},
  {Vargas-Gonzalez}, {Vargas-Magana}, {Vazquez-Mata}, {Vivek}, {Wake}, {Wang},
  {Weaver}, {Weijmans}, {Wild}, {Wilson}, {Wilson}, {Wolthuis}, {Wood-Vasey},
  {Yan}, {Yang}, {Yeche}, {Zamora}, {Zarrouk}, {Zasowski}, {Zhang}, {Zhao},
  {Zhao}, {Zheng}, {Zheng}, {Zhu}, \& {Zou}}]{APOGEE_Dr16}
{Ahumada}, R., {Allende Prieto}, C., {Almeida}, A., {et~al.} 2019, arXiv
  e-prints, arXiv:1912.02905

\bibitem[{{Ahumada} {et~al.}(2020){Ahumada}, {Prieto}, {Almeida}, {Anders},
  {Anderson}, {Andrews}, {Anguiano}, {Arcodia}, {Armengaud}, {Aubert}, {Avila},
  {Avila-Reese}, {Badenes}, {Balland}, {Barger}, {Barrera-Ballesteros}, {Basu},
  {Bautista}, {Beaton}, {Beers}, {Benavides}, {Bender}, {Bernardi}, {Bershady},
  {Beutler}, {Bidin}, {Bird}, {Bizyaev}, {Blanc}, {Blanton}, {Boquien},
  {Borissova}, {Bovy}, {Brandt}, {Brinkmann}, {Brownstein}, {Bundy}, {Bureau},
  {Burgasser}, {Burtin}, {Cano-D{\'\i}az}, {Capasso}, {Cappellari}, {Carrera},
  {Chabanier}, {Chaplin}, {Chapman}, {Cherinka}, {Chiappini}, {Doohyun Choi},
  {Chojnowski}, {Chung}, {Clerc}, {Coffey}, {Comerford}, {Comparat}, {da
  Costa}, {Cousinou}, {Covey}, {Crane}, {Cunha}, {Ilha}, {Dai}, {Damsted},
  {Darling}, {Davidson}, {Davies}, {Dawson}, {De}, {de la Macorra}, {De Lee},
  {Queiroz}, {Deconto Machado}, {de la Torre}, {Dell'Agli}, {du Mas des
  Bourboux}, {Diamond-Stanic}, {Dillon}, {Donor}, {Drory}, {Duckworth},
  {Dwelly}, {Ebelke}, {Eftekharzadeh}, {Davis Eigenbrot}, {Elsworth},
  {Eracleous}, {Erfanianfar}, {Escoffier}, {Fan}, {Farr},
  {Fern{\'a}ndez-Trincado}, {Feuillet}, {Finoguenov}, {Fofie},
  {Fraser-McKelvie}, {Frinchaboy}, {Fromenteau}, {Fu}, {Galbany}, {Garcia},
  {Garc{\'\i}a-Hern{\'a}ndez}, {Oehmichen}, {Ge}, {Maia}, {Geisler}, {Gelfand},
  {Goddy}, {Gonzalez-Perez}, {Grabowski}, {Green}, {Grier}, {Guo}, {Guy},
  {Harding}, {Hasselquist}, {Hawken}, {Hayes}, {Hearty}, {Hekker}, {Hogg},
  {Holtzman}, {Horta}, {Hou}, {Hsieh}, {Huber}, {Hunt}, {Chitham}, {Imig},
  {Jaber}, {Angel}, {Johnson}, {Jones}, {J{\"o}nsson}, {Jullo}, {Kim},
  {Kinemuchi}, {Kirkpatrick}, {Kite}, {Klaene}, {Kneib}, {Kollmeier}, {Kong},
  {Kounkel}, {Krishnarao}, {Lacerna}, {Lan}, {Lane}, {Law}, {Le Goff}, {Leung},
  {Lewis}, {Li}, {Lian}, {Lin}, {Long}, {Longa-Pe{\~n}a}, {Lundgren}, {Lyke},
  {Ted Mackereth}, {MacLeod}, {Majewski}, {Manchado}, {Maraston}, {Martini},
  {Masseron}, {Masters}, {Mathur}, {McDermid}, {Merloni}, {Merrifield},
  {M{\'e}sz{\'a}ros}, {Miglio}, {Minniti}, {Minsley}, {Miyaji}, {Mohammad},
  {Mosser}, {Mueller}, {Muna}, {Mu{\~n}oz-Guti{\'e}rrez}, {Myers}, {Nadathur},
  {Nair}, {Nandra}, {do Nascimento}, {Nevin}, {Newman}, {Nidever}, {Nitschelm},
  {Noterdaeme}, {O'Connell}, {Olmstead}, {Oravetz}, {Oravetz}, {Osorio},
  {Pace}, {Padilla}, {Palanque-Delabrouille}, {Palicio}, {Pan}, {Pan},
  {Parker}, {Paviot}, {Peirani}, {Ram{\'r}ez}, {Penny}, {Percival},
  {Perez-Fournon}, {P{\'e}rez-R{\`a}fols}, {Petitjean}, {Pieri},
  {Pinsonneault}, {Poovelil}, {Povick}, {Prakash}, {Price-Whelan}, {Raddick},
  {Raichoor}, {Ray}, {Rembold}, {Rezaie}, {Riffel}, {Riffel}, {Rix}, {Robin},
  {Roman-Lopes}, {Rom{\'a}n-Z{\'u}{\~n}iga}, {Rose}, {Ross}, {Rossi},
  {Rowlands}, {Rubin}, {Salvato}, {S{\'a}nchez}, {S{\'a}nchez-Menguiano},
  {S{\'a}nchez-Gallego}, {Sayres}, {Schaefer}, {Schiavon}, {Schimoia},
  {Schlafly}, {Schlegel}, {Schneider}, {Schultheis}, {Schwope}, {Seo},
  {Serenelli}, {Shafieloo}, {Shamsi}, {Shao}, {Shen}, {Shetrone}, {Shirley},
  {Aguirre}, {Simon}, {Skrutskie}, {Slosar}, {Smethurst}, {Sobeck}, {Sodi},
  {Souto}, {Stark}, {Stassun}, {Steinmetz}, {Stello}, {Stermer},
  {Storchi-Bergmann}, {Streblyanska}, {Stringfellow}, {Stutz}, {Su{\'a}rez},
  {Sun}, {Taghizadeh-Popp}, {Talbot}, {Tayar}, {Thakar}, {Theriault}, {Thomas},
  {Thomas}, {Tinker}, {Tojeiro}, {Toledo}, {Tremonti}, {Troup}, {Tuttle},
  {Unda-Sanzana}, {Valentini}, {Vargas-Gonz{\'a}lez}, {Vargas-Maga{\~n}a},
  {V{\'a}zquez-Mata}, {Vivek}, {Wake}, {Wang}, {Weaver}, {Weijmans}, {Wild},
  {Wilson}, {Wilson}, {Wolthuis}, {Wood-Vasey}, {Yan}, {Yang}, {Y{\`e}che},
  {Zamora}, {Zarrouk}, {Zasowski}, {Zhang}, {Zhao}, {Zhao}, {Zheng}, {Zheng},
  {Zhu}, \& {Zou}}]{2020ApJS..249....3A}
{Ahumada}, R., {Prieto}, C.~A., {Almeida}, A., {et~al.} 2020, \apjs, 249, 3

\bibitem[{Alonso-Floriano {et~al.}(2015)Alonso-Floriano, Morales, Caballero,
  Montes, Klutsch, Mundt, Cort{\'e}s-Contreras, Ribas, Reiners, Amado,
  {et~al.}}]{alonso2015carmenes}
Alonso-Floriano, F., Morales, J., Caballero, J., {et~al.} 2015, Astronomy \&
  Astrophysics, 577, A128

\bibitem[{{Alvarez} \& {Plez}(1998)}]{plez1998}
{Alvarez}, R. \& {Plez}, B. 1998, 330, 1109

\bibitem[{Artigau {et~al.}(2014)Artigau, Kouach, Donati, Doyon, Delfosse,
  Baratchart, Lacombe, Moutou, Rabou, Par{\`e}s, {et~al.}}]{artigau2014spirou}
Artigau, {\'E}., Kouach, D., Donati, J.-F., {et~al.} 2014, in Ground-based and
  Airborne Instrumentation for Astronomy V, Vol. 9147, International Society
  for Optics and Photonics, 914715

\bibitem[{Barber {et~al.}(2006)Barber, Tennyson, Harris, \&
  Tolchenov}]{barber2006high}
Barber, R., Tennyson, J., Harris, G.~J., \& Tolchenov, R. 2006, Monthly Notices
  of the Royal Astronomical Society, 368, 1087

\bibitem[{{Blanco-Cuaresma}(2019)}]{ispec2019sbc}
{Blanco-Cuaresma}, S. 2019, mnras, 486, 2075

\bibitem[{Blanco-Cuaresma {et~al.}(2014)Blanco-Cuaresma, Soubiran, Heiter, \&
  Jofr{\'e}}]{blanco2014determining}
Blanco-Cuaresma, S., Soubiran, C., Heiter, U., \& Jofr{\'e}, P. 2014, Astronomy
  \& Astrophysics, 569, A111

\bibitem[{Bochanski {et~al.}(2007)Bochanski, Munn, Hawley, West, Covey, \&
  Schneider}]{bochanski2007exploring}
Bochanski, J.~J., Munn, J.~A., Hawley, S.~L., {et~al.} 2007, The Astronomical
  Journal, 134, 2418

\bibitem[{Bonfils {et~al.}(2013)Bonfils, Delfosse, Udry, Forveille, Mayor,
  Perrier, Bouchy, Gillon, Lovis, Pepe, {et~al.}}]{bonfils2013harps}
Bonfils, X., Delfosse, X., Udry, S., {et~al.} 2013, Astronomy \& Astrophysics,
  549, A109

\bibitem[{Bowen \& Vaughan(1973)}]{bowen1973optical}
Bowen, I. \& Vaughan, A. 1973, Applied Optics, 12, 1430

\bibitem[{{Boyajian} {et~al.}(2013){Boyajian}, {von Braun}, {van Belle},
  {Farrington}, {Schaefer}, {Jones}, {White}, {McAlister}, {ten Brummelaar},
  {Ridgway}, {Gies}, {Sturmann}, {Sturmann}, {Turner}, {Goldfinger}, \&
  {Vargas}}]{Boayajian2013}
{Boyajian}, T.~S., {von Braun}, K., {van Belle}, G., {et~al.} 2013, apj, 771,
  40

\bibitem[{Bressan {et~al.}(2012)Bressan, Marigo, Girardi, Salasnich, Dal~Cero,
  Rubele, \& Nanni}]{bressan2012parsec}
Bressan, A., Marigo, P., Girardi, L., {et~al.} 2012, Monthly Notices of the
  Royal Astronomical Society, 427, 127

\bibitem[{Casagrande {et~al.}(2008)Casagrande, Flynn, \&
  Bessell}]{casagrande2008m}
Casagrande, L., Flynn, C., \& Bessell, M. 2008, Monthly Notices of the Royal
  Astronomical Society, 389, 585

\bibitem[{Delfosse {et~al.}(2000)Delfosse, Forveille, S{\'e}gransan, Beuzit,
  Udry, Perrier, \& Mayor}]{delfosse2000accurate}
Delfosse, X., Forveille, T., S{\'e}gransan, D., {et~al.} 2000, arXiv preprint
  astro-ph/0010586

\bibitem[{Deshpande {et~al.}(2013)Deshpande, Blake, Bender, Mahadevan, Terrien,
  Carlberg, Zasowski, Crepp, Rajpurohit, Reyl{\'e},
  {et~al.}}]{deshpande2013sdss}
Deshpande, R., Blake, C., Bender, C., {et~al.} 2013, The Astronomical Journal,
  146, 156

\bibitem[{Dorn {et~al.}(2014)Dorn, Anglada-Escude, Baade, Bristow, Follert,
  Gojak, Grunhut, Hatzes, Heiter, Hilker, {et~al.}}]{dorn2014crires+}
Dorn, R.~J., Anglada-Escude, G., Baade, D.~a., {et~al.} 2014, The Messenger,
  156

\bibitem[{France {et~al.}(2013)France, Froning, Linsky, Roberge, Stocke, Tian,
  Bushinsky, D{\'e}sert, Mauas, Vieytes, {et~al.}}]{france2013ultraviolet}
France, K., Froning, C.~S., Linsky, J.~L., {et~al.} 2013, The Astrophysical
  Journal, 763, 149

\bibitem[{Gaidos {et~al.}(2014)Gaidos, Mann, L{\'e}pine, Buccino, James,
  Ansdell, Petrucci, Mauas, \& Hilton}]{gaidos2014trumpeting}
Gaidos, E., Mann, A., L{\'e}pine, S., {et~al.} 2014, Monthly Notices of the
  Royal Astronomical Society, 443, 2561

\bibitem[{Garcia~P{\'e}rez {et~al.}(2016)Garcia~P{\'e}rez, Prieto, Holtzman,
  Shetrone, M{\'e}sz{\'a}ros, Bizyaev, Carrera, Cunha,
  Garc{\'\i}a-Hern{\'a}ndez, Johnson, {et~al.}}]{perez2016aspcap}
Garcia~P{\'e}rez, A.~E., Prieto, C.~A., Holtzman, J.~A., {et~al.} 2016, The
  Astronomical Journal, 151, 144

\bibitem[{Gunn {et~al.}(2006)Gunn, Siegmund, Mannery, Owen, Hull, Leger, Carey,
  Knapp, York, Boroski, {et~al.}}]{gunn20062}
Gunn, J.~E., Siegmund, W.~A., Mannery, E.~J., {et~al.} 2006, The Astronomical
  Journal, 131, 2332

\bibitem[{{Gustafsson} {et~al.}(2008){Gustafsson}, {Edvardsson}, {Eriksson},
  {J{\o}rgensen}, {Nordlund}, \& {Plez}}]{MARCS}
{Gustafsson}, B., {Edvardsson}, B., {Eriksson}, K., {et~al.} 2008, aap, 486,
  951

\bibitem[{{Hejazi} {et~al.}(2019){Hejazi}, {Lepine}, {Homeier}, {Rich}, \&
  {Shara}}]{Hejazi2019Chemical}
{Hejazi}, N., {Lepine}, S., {Homeier}, D., {Rich}, R.~M., \& {Shara}, M.~M.
  2019, arXiv e-prints, arXiv:1911.04612

\bibitem[{Henry {et~al.}(2006)Henry, Jao, Subasavage, Beaulieu, Ianna, Costa,
  \& M{\'e}ndez}]{henry2006solar}
Henry, T.~J., Jao, W.-C., Subasavage, J.~P., {et~al.} 2006, The Astronomical
  Journal, 132, 2360

\bibitem[{Holmberg {et~al.}(2007)Holmberg, Nordstr{\"o}m, \&
  Andersen}]{holmberg2007geneva}
Holmberg, J., Nordstr{\"o}m, B., \& Andersen, J. 2007, Astronomy \&
  Astrophysics, 475, 519

\bibitem[{Holtzman {et~al.}(2018)Holtzman, Hasselquist, Shetrone, Cunha,
  Prieto, Anguiano, Bizyaev, Bovy, Casey, Edvardsson,
  {et~al.}}]{holtzman2018apogee}
Holtzman, J.~A., Hasselquist, S., Shetrone, M., {et~al.} 2018, The Astronomical
  Journal, 156, 125

\bibitem[{Irwin {et~al.}(2014)Irwin, Berta-Thompson, Charbonneau, Dittmann,
  Falco, Newton, \& Nutzman}]{irwin2014mearth}
Irwin, J.~M., Berta-Thompson, Z.~K., Charbonneau, D., {et~al.} 2014, arXiv
  preprint arXiv:1409.0891

\bibitem[{{J{\"o}nsson} {et~al.}(2020){J{\"o}nsson}, {Holtzman}, {Allende
  Prieto}, {Cunha}, {Garc{\'\i}a-Hern{\'a}ndez}, {Hasselquist}, {Masseron},
  {Osorio}, {Shetrone}, {Smith}, {Stringfellow}, {Bizyaev}, {Edvardsson},
  {Majewski}, {M{\'e}sz{\'a}ros}, {Souto}, {Zamora}, {Beaton}, {Bovy}, {Donor},
  {Pinsonneault}, {Poovelil}, \& {Sobeck}}]{2020AJ....160..120J}
{J{\"o}nsson}, H., {Holtzman}, J.~A., {Allende Prieto}, C., {et~al.} 2020, \aj,
  160, 120

\bibitem[{Latham {et~al.}(2005)Latham, Brown, Monet, Everett, Esquerdo, \&
  Hergenrother}]{latham2005kepler}
Latham, D., Brown, T., Monet, D., {et~al.} 2005, in Bulletin of the American
  Astronomical Society, Vol.~37, 1340

\bibitem[{L{\'e}pine \& Gaidos(2011)}]{lepine2011all}
L{\'e}pine, S. \& Gaidos, E. 2011, The Astronomical Journal, 142, 138

\bibitem[{L{\'e}pine {et~al.}(2007)L{\'e}pine, Rich, \&
  Shara}]{lepine2007revised}
L{\'e}pine, S., Rich, R.~M., \& Shara, M.~M. 2007, The Astrophysical Journal,
  669, 1235

\bibitem[{L{\'e}pine \& Shara(2005)}]{lepine2005catalog}
L{\'e}pine, S. \& Shara, M.~M. 2005, The Astronomical Journal, 129, 1483

\bibitem[{L{\'o}pez-Valdivia {et~al.}(2019)L{\'o}pez-Valdivia, Mace, Sokal,
  Hussaini, Kidder, Mann, Gosnell, Oh, Kesseli, Muirhead,
  {et~al.}}]{lopez2019effective}
L{\'o}pez-Valdivia, R., Mace, G.~N., Sokal, K.~R., {et~al.} 2019, arXiv
  preprint arXiv:1905.05076

\bibitem[{Majewski {et~al.}(2017)Majewski, Schiavon, Frinchaboy, Prieto,
  Barkhouser, Bizyaev, Blank, Brunner, Burton, Carrera,
  {et~al.}}]{majewski2017apache}
Majewski, S.~R., Schiavon, R.~P., Frinchaboy, P.~M., {et~al.} 2017, The
  Astronomical Journal, 154, 94

\bibitem[{Mann {et~al.}(2013)Mann, Brewer, Gaidos, L{\'e}pine, \&
  Hilton}]{mann2013prospecting}
Mann, A.~W., Brewer, J.~M., Gaidos, E., L{\'e}pine, S., \& Hilton, E.~J. 2013,
  The Astronomical Journal, 145, 52

\bibitem[{Mann {et~al.}(2015)Mann, Feiden, Gaidos, Boyajian, \& von
  Braun}]{mann2015constrain}
Mann, A.~W., Feiden, G.~A., Gaidos, E., Boyajian, T., \& von Braun, K. 2015,
  The Astrophysical Journal, 804, 64

\bibitem[{Markwardt(2009)}]{markwardt2009non}
Markwardt, C.~B. 2009, arXiv preprint arXiv:0902.2850

\bibitem[{Ness {et~al.}(2016)Ness, Hogg, Rix, Martig, Pinsonneault, \&
  Ho}]{ness2016spectroscopic}
Ness, M., Hogg, D.~W., Rix, H.-W., {et~al.} 2016, The Astrophysical Journal,
  823, 114

\bibitem[{{\"O}nehag {et~al.}(2012){\"O}nehag, Heiter, Gustafsson, Piskunov,
  Plez, \& Reiners}]{onehag2012m}
{\"O}nehag, A., Heiter, U., Gustafsson, B., {et~al.} 2012, Astronomy \&
  Astrophysics, 542, A33

\bibitem[{Origlia {et~al.}(2014)Origlia, Oliva, Baffa, Falcini, Giani, Massi,
  Montegriffo, Sanna, Scuderi, Sozzi, {et~al.}}]{origlia2014high}
Origlia, L., Oliva, E., Baffa, C., {et~al.} 2014, in Ground-based and Airborne
  Instrumentation for Astronomy V, Vol. 9147, International Society for Optics
  and Photonics, 91471E

\bibitem[{Passegger {et~al.}(2019)Passegger, Schweitzer, Shulyak, Nagel,
  Hauschildt, Reiners, Amado, Caballero, Cort{\'e}s-Contreras,
  Dom{\'\i}nguez-Fern{\'a}ndez, {et~al.}}]{passegger2019carmenes}
Passegger, V., Schweitzer, A., Shulyak, D., {et~al.} 2019, Astronomy \&
  Astrophysics, 627, A161

\bibitem[{Piskunov {et~al.}(1995)Piskunov, Kupka, Ryabchikova, Weiss, \&
  Jeffery}]{piskunov1995vald}
Piskunov, N., Kupka, F., Ryabchikova, T., Weiss, W., \& Jeffery, C. 1995,
  Astronomy and Astrophysics Supplement Series, 112, 525

\bibitem[{Plez(2012)}]{plez2012turbospectrum}
Plez, B. 2012, Astrophysics Source Code Library, 1, 05004

\bibitem[{Polyansky {et~al.}(2018)Polyansky, Kyuberis, Zobov, Tennyson,
  Yurchenko, \& Lodi}]{polyansky2018exomol}
Polyansky, O.~L., Kyuberis, A.~A., Zobov, N.~F., {et~al.} 2018, Monthly Notices
  of the Royal Astronomical Society, 480, 2597

\bibitem[{Quirrenbach {et~al.}(2014)Quirrenbach, Amado, Caballero, Mundt,
  Reiners, Ribas, Seifert, Abril, Aceituno, Alonso-Floriano,
  {et~al.}}]{quirrenbach2014carmenes}
Quirrenbach, A., Amado, P., Caballero, J., {et~al.} 2014, in Ground-based and
  airborne instrumentation for astronomy V, Vol. 9147, International Society
  for Optics and Photonics, 91471F

\bibitem[{Rajpurohit {et~al.}(2018{\natexlab{a}})Rajpurohit, Allard,
  Rajpurohit, Sharma, Teixeira, Mousis, \& Kamlesh}]{rajpurohit2018carmenes}
Rajpurohit, A., Allard, F., Rajpurohit, S., {et~al.} 2018{\natexlab{a}},
  Astronomy \& Astrophysics, 620, A180

\bibitem[{Rajpurohit {et~al.}(2018{\natexlab{b}})Rajpurohit, Allard, Teixeira,
  Homeier, Rajpurohit, \& Mousis}]{rajpurohit2018apogee}
Rajpurohit, A., Allard, F., Teixeira, G., {et~al.} 2018{\natexlab{b}},
  Astronomy \& Astrophysics, 610, A19

\bibitem[{Rojas-Ayala {et~al.}(2010)Rojas-Ayala, Covey, Muirhead, \&
  Lloyd}]{rojas2010metal}
Rojas-Ayala, B., Covey, K.~R., Muirhead, P.~S., \& Lloyd, J.~P. 2010, The
  Astrophysical Journal Letters, 720, L113

\bibitem[{Rojas-Ayala {et~al.}(2012)Rojas-Ayala, Covey, Muirhead, \&
  Lloyd}]{rojas2012metallicity}
Rojas-Ayala, B., Covey, K.~R., Muirhead, P.~S., \& Lloyd, J.~P. 2012, The
  Astrophysical Journal, 748, 93

\bibitem[{Santos {et~al.}(2013)Santos, Sousa, Mortier, Neves, Adibekyan,
  Tsantaki, Mena, Bonfils, Israelian, Mayor, {et~al.}}]{santos2013sweet}
Santos, N., Sousa, S., Mortier, A., {et~al.} 2013, Astronomy \& Astrophysics,
  556, A150

\bibitem[{Sarmento {et~al.}(2020)Sarmento, Mena, Rojas-Ayala, \&
  Blanco-Cuaresma}]{sarmento2020derivation}
Sarmento, P., Mena, E.~D., Rojas-Ayala, B., \& Blanco-Cuaresma, S. 2020, arXiv
  preprint arXiv:2001.01995

\bibitem[{Scalo {et~al.}(2007)Scalo, Kaltenegger, Segura, Fridlund, Ribas,
  Kulikov, Grenfell, Rauer, Odert, Leitzinger, {et~al.}}]{scalo2007m}
Scalo, J., Kaltenegger, L., Segura, A., {et~al.} 2007, Astrobiology, 7, 85

\bibitem[{Schmidt {et~al.}(2016)Schmidt, Wagoner, Johnson, Davenport, Stassun,
  Souto, \& Ge}]{schmidt2016examining}
Schmidt, S.~J., Wagoner, E.~L., Johnson, J.~A., {et~al.} 2016, Monthly Notices
  of the Royal Astronomical Society, 460, 2611

\bibitem[{Segura {et~al.}(2005)Segura, Kasting, Meadows, Cohen, Scalo, Crisp,
  Butler, \& Tinetti}]{segura2005biosignatures}
Segura, A., Kasting, J.~F., Meadows, V., {et~al.} 2005, Astrobiology, 5, 706

\bibitem[{Shetrone {et~al.}(2015)Shetrone, Bizyaev, Lawler, Prieto, Johnson,
  Smith, Cunha, Holtzman, P{\'e}rez, M{\'e}sz{\'a}ros,
  {et~al.}}]{shetrone2015sdss}
Shetrone, M., Bizyaev, D., Lawler, J.~E., {et~al.} 2015, The Astrophysical
  Journal Supplement Series, 221, 24

\bibitem[{Sousa {et~al.}(2011)Sousa, Santos, Israelian, Lovis, Mayor, Silva, \&
  Udry}]{Sousa2011a}
Sousa, S.~G., Santos, N.~C., Israelian, G., {et~al.} 2011, Astronomy \&
  Astrophysics, 526, A99

\bibitem[{{Souto} {et~al.}(2017){Souto}, {Cunha},
  {Garc{\'{\i}}a-Hern{\'a}ndez}, {Zamora}, {Allende Prieto}, {Smith},
  {Mahadevan}, {Blake}, {Johnson}, {J{\"o}nsson}, {Pinsonneault}, {Holtzman},
  {Majewski}, {Shetrone}, {Teske}, {Nidever}, {Schiavon}, {Sobeck},
  {Garc{\'{\i}}a P{\'e}rez}, {G{\'o}mez Maqueo Chew}, \& {Stassun}}]{souto2017}
{Souto}, D., {Cunha}, K., {Garc{\'{\i}}a-Hern{\'a}ndez}, D.~A., {et~al.} 2017,
  835, 239

\bibitem[{{Souto} {et~al.}(2020){Souto}, {Cunha}, {Smith}, {Allende Prieto},
  {Burgasser}, {Covey}, {Garcia-Hernandez}, {Holtzman}, {Johnson}, {Jonsson},
  {Mahadevan}, {Majewski}, {Masseron}, {Shetrone}, {Rojas-Ayala}, {Sobeck},
  {Stassun}, {Terrien}, {Teske}, {Wanderley}, \& {Zamora}}]{Souto2020}
{Souto}, D., {Cunha}, K., {Smith}, V.~V., {et~al.} 2020, arXiv e-prints,
  arXiv:2001.05597

\bibitem[{Souto {et~al.}(2018)Souto, Unterborn, Smith, Cunha, Teske, Covey,
  Rojas-Ayala, Garc{\'\i}a-Hern{\'a}ndez, Stassun, Zamora,
  {et~al.}}]{souto2018stellar}
Souto, D., Unterborn, C.~T., Smith, V.~V., {et~al.} 2018, The Astrophysical
  Journal Letters, 860, L15

\bibitem[{Suda {et~al.}(2008)Suda, Katsuta, Yamada, Suwa, Ishizuka, Komiya,
  Sorai, Aikawa, \& Fujimoto}]{suda2008stellar}
Suda, T., Katsuta, Y., Yamada, S., {et~al.} 2008, Publications of the
  Astronomical Society of Japan, 60, 1159

\bibitem[{Terrien {et~al.}(2015)Terrien, Mahadevan, Deshpande, \&
  Bender}]{terrien2015near}
Terrien, R.~C., Mahadevan, S., Deshpande, R., \& Bender, C.~F. 2015, The
  Astrophysical Journal Supplement Series, 220, 16

\bibitem[{Veyette {et~al.}(2017)Veyette, Muirhead, Mann, Brewer, Allard, \&
  Homeier}]{veyette2017physically}
Veyette, M.~J., Muirhead, P.~S., Mann, A.~W., {et~al.} 2017, The Astrophysical
  Journal, 851, 26

\bibitem[{Wildi {et~al.}(2017)Wildi, Blind, Reshetov, Hernandez, Genolet,
  Conod, Sordet, Segovilla, Rasilla, Brousseau, {et~al.}}]{wildi2017nirps}
Wildi, F., Blind, N., Reshetov, V., {et~al.} 2017, in Techniques and
  Instrumentation for Detection of Exoplanets VIII, Vol. 10400, International
  Society for Optics and Photonics, 1040018

\bibitem[{Zasowski {et~al.}(2013)Zasowski, Johnson, Frinchaboy, Majewski,
  Nidever, Pinto, Girardi, Andrews, Chojnowski, Cudworth,
  {et~al.}}]{zasowski2013target}
Zasowski, G., Johnson, J.~A., Frinchaboy, P., {et~al.} 2013, The Astronomical
  Journal, 146, 81

\end{thebibliography}

\begin{appendix}

\section{Parameter combination list for the normalization}

\tablehead{\hline\hline $T_\mathrm{eff}$ (K) & $\log g$ (dex) & $[M/H]$ (dex) \\}
\label{table:parameter_normalization}
\centering  
\bottomcaption{Parameter combination list used for the template normalization. Values taken from PARSEC isochrones \citep{bressan2012parsec} for stars with ages between $10^9$ and $10^{10}$ years, and rounded to the nearest multiple of $100\,$K (for $T_\mathrm{eff}$) or $0.1\,$dex (for $\log g$ and metallicity).}

\begin{supertabular}{ccc}
\hline
2500	&	-0.4	&	5.3	\\
2500	&	0	&	5.2	\\
2500	&	0.2	&	5.1	\\
2500	&	0.4	&	5.1	\\
2600	&	-0.6	&	5.3	\\
2600	&	-0.4	&	5.2	\\
2600	&	-0.2	&	5.2	\\
2600	&	0	&	5.1	\\
2600	&	0	&	5.2	\\
2600	&	0.2	&	5.1	\\
2600	&	0.4	&	5.1	\\
2600	&	0.4	&	5	\\
2700	&	-0.8	&	5.3	\\
2700	&	-0.6	&	5.3	\\
2700	&	-0.4	&	5.2	\\
2700	&	-0.2	&	5.2	\\
2700	&	0	&	5.1	\\
2700	&	0.2	&	5.1	\\
2700	&	0.4	&	5	\\
2800	&	-1	&	5.3	\\
2800	&	-1	&	5.4	\\
2800	&	-0.8	&	5.3	\\
2800	&	-0.6	&	5.3	\\
2800	&	-0.2	&	5.2	\\
2800	&	0	&	5.1	\\
2800	&	0.2	&	5.1	\\
2800	&	0.4	&	5	\\
2900	&	-1	&	5.3	\\
2900	&	-0.8	&	5.3	\\
2900	&	-0.4	&	5.2	\\
2900	&	-0.2	&	5.1	\\
2900	&	0	&	5.1	\\
2900	&	0.2	&	5	\\
2900	&	0.4	&	5	\\
2900	&	0.4	&	4.9	\\
3000	&	-1	&	5.3	\\
3000	&	-0.8	&	5.2	\\
3000	&	-0.8	&	5.3	\\
3000	&	-0.6	&	5.2	\\
3000	&	-0.4	&	5.1	\\
3000	&	-0.2	&	5.1	\\
3000	&	0	&	5	\\
3000	&	0.2	&	5	\\
3000	&	0.4	&	4.9	\\
3100	&	-1	&	5.3	\\
3100	&	-0.8	&	5.2	\\
3100	&	-0.6	&	5.2	\\
3100	&	-0.4	&	5.1	\\
3100	&	-0.2	&	5.1	\\
3100	&	-0.2	&	5	\\
3100	&	0	&	5	\\
3100	&	0.2	&	4.9	\\
3100	&	0.4	&	4.9	\\
3100	&	0.4	&	4.8	\\
3200	&	-1	&	5.2	\\
3200	&	-1	&	5.3	\\
3200	&	-0.8	&	5.2	\\
3200	&	-0.6	&	5.1	\\
3200	&	-0.4	&	5.1	\\
3200	&	-0.2	&	5	\\
3200	&	0	&	4.9	\\
3200	&	0.2	&	4.9	\\
3200	&	0.2	&	4.8	\\
3200	&	0.4	&	4.8	\\
3300	&	-0.8	&	5.1	\\
3300	&	-0.6	&	5.1	\\
3300	&	-0.4	&	5	\\
3300	&	-0.2	&	4.9	\\
3300	&	-0.2	&	5	\\
3300	&	0	&	4.9	\\
3300	&	0	&	4.8	\\
3300	&	0.2	&	4.8	\\
3300	&	0.4	&	4.8	\\
3400	&	-1	&	5.2	\\
3400	&	-0.8	&	5.1	\\
3400	&	-0.6	&	5.1	\\
3400	&	-0.6	&	5	\\
3400	&	-0.4	&	5	\\
3400	&	-0.2	&	4.9	\\
3400	&	0	&	4.8	\\
3400	&	0.2	&	4.8	\\
3400	&	0.2	&	4.7	\\
3400	&	0.4	&	4.8	\\
3400	&	0.4	&	4.7	\\
3500	&	-1	&	5.2	\\
3500	&	-1	&	5.1	\\
3500	&	-0.8	&	5.1	\\
3500	&	-0.6	&	5	\\
3500	&	-0.4	&	5	\\
3500	&	-0.4	&	4.9	\\
3500	&	-0.2	&	4.9	\\
3500	&	-0.2	&	4.8	\\
3500	&	0.2	&	4.8	\\
3500	&	0.2	&	4.7	\\
3500	&	0.4	&	4.7	\\
3600	&	-1	&	5.1	\\
3600	&	-0.8	&	5.1	\\
3600	&	-0.8	&	5	\\
3600	&	-0.6	&	5	\\
3600	&	-0.6	&	4.9	\\
3600	&	-0.4	&	4.9	\\
3600	&	-0.4	&	4.8	\\
3600	&	-0.2	&	4.8	\\
3600	&	0	&	4.8	\\
3600	&	0	&	4.7	\\
3600	&	0.2	&	4.7	\\
3600	&	0.4	&	4.7	\\
3700	&	-1	&	5.1	\\
3700	&	-1	&	5	\\
3700	&	-0.8	&	5	\\
3700	&	-0.8	&	4.9	\\
3700	&	-0.6	&	4.9	\\
3700	&	-0.4	&	4.8	\\
3700	&	-0.2	&	4.8	\\
3700	&	-0.2	&	4.7	\\
3700	&	0	&	4.7	\\
3700	&	0.2	&	4.7	\\
3700	&	0.4	&	4.7	\\
3700	&	0.4	&	4.6	\\
3800	&	-1	&	5	\\
3800	&	-1	&	4.9	\\
3800	&	-0.8	&	4.9	\\
3800	&	-0.6	&	4.9	\\
3800	&	-0.6	&	4.8	\\
3800	&	-0.4	&	4.8	\\
3800	&	0	&	4.7	\\
3800	&	0.2	&	4.7	\\
3800	&	0.2	&	4.6	\\
3900	&	-1	&	4.9	\\
3900	&	-0.8	&	4.9	\\
3900	&	-0.8	&	4.8	\\
3900	&	-0.4	&	4.8	\\
3900	&	-0.4	&	4.7	\\
3900	&	-0.2	&	4.7	\\
3900	&	0	&	4.7	\\
3900	&	0.4	&	4.7	\\
3900	&	0.4	&	4.6	\\
4000	&	-0.8	&	4.8	\\
4000	&	-0.6	&	4.8	\\
4000	&	-0.4	&	4.7	\\
4000	&	0	&	4.7	\\
4000	&	0.2	&	4.7	\\
4000	&	0.2	&	4.6	\\
4000	&	0.4	&	4.7	\\
\hline
\end{supertabular}

\section{More example synthetic and observed spectra comparisons}

\begin{landscape}

\begin{figure}
\resizebox{\hsize}{!}{\includegraphics{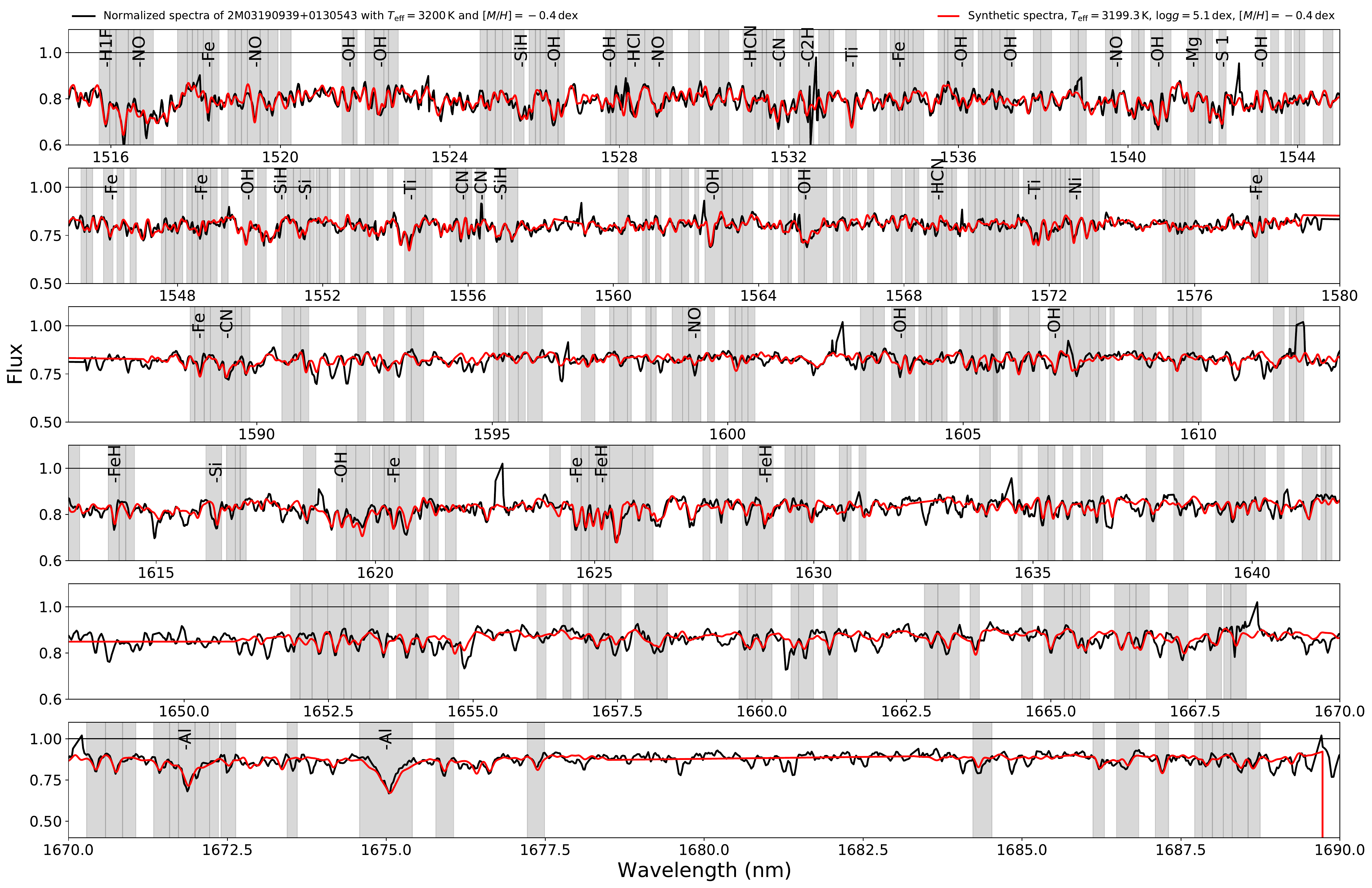}}
\caption{Comparison between APOGEE spectra of the star 2M03190939+0130543 (black, normalized using a synthetic spectra with $T_\mathrm{eff} = 3200\,$K, $\log g = 5.1$\,dex and $[M/H] = -0.4\,$dex) and the best fitting synthetic spectra (red, straight line) for APOGEE wavelength range. In gray highlight are the areas used for $\chi ^2 $ minimization by our pipeline's algorithm.
The best fitting parameters derived were $T_\mathrm{eff}=3199\pm100$\,K,$\log g=5.1\pm0.2$\,dex, $[M/H] = -0.37\pm0.1$\,dex. The available ASPCAP parameters are $T_\mathrm{eff}=3287\pm60$\,K, $\log g=5.17\pm0.14$\,dex and $[M/H] = -0.57\pm0.02$\,dex.}
\label{Mdwarf_4}
\end{figure}

\end{landscape}

\begin{landscape}

\begin{figure}
\resizebox{\hsize}{!}{\includegraphics{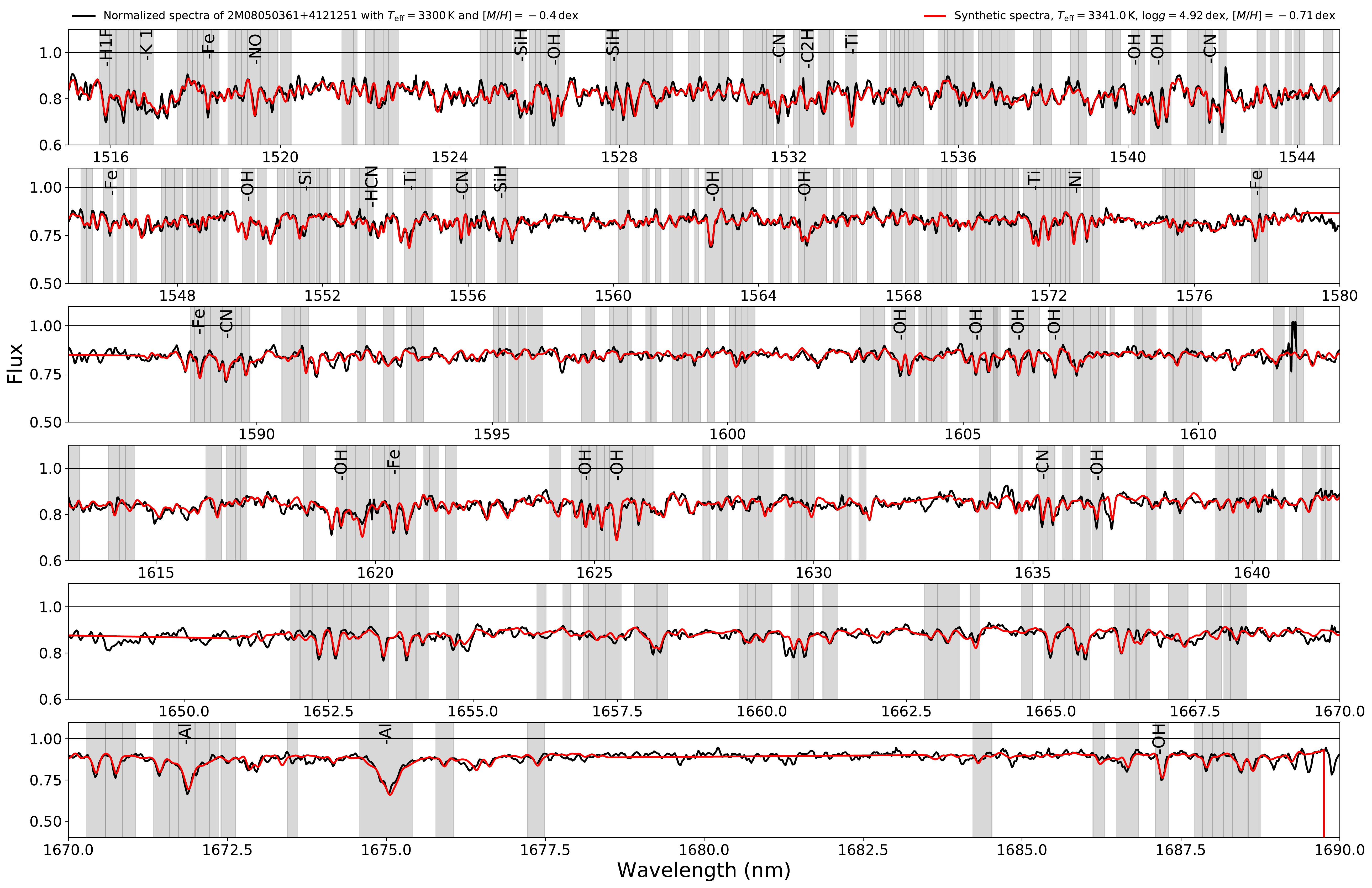}}
\caption{Comparison between APOGEE spectra of the star 2M08050361+4121251 (G 111-52, black, normalized using a synthetic spectra with $T_\mathrm{eff} = 3400\,$K, $\log g = 5.0$\,dex and $[M/H]=-0.6\,$dex) and the best fitting synthetic spectra (red, straight line) for APOGEE wavelength range.
In gray highlight are the areas used for $\chi ^2 $ minimization by our pipeline's algorithm.
The best fitting parameters derived were $T_\mathrm{eff}=3413\pm100$\,K,$\log g=4.98\pm0.2$\,dex, $[M/H] = -0.78\pm0.1$\,dex.}
\label{Mdwarf_1}
\end{figure}

\end{landscape}

\begin{landscape}

\begin{figure}
\resizebox{\hsize}{!}{\includegraphics{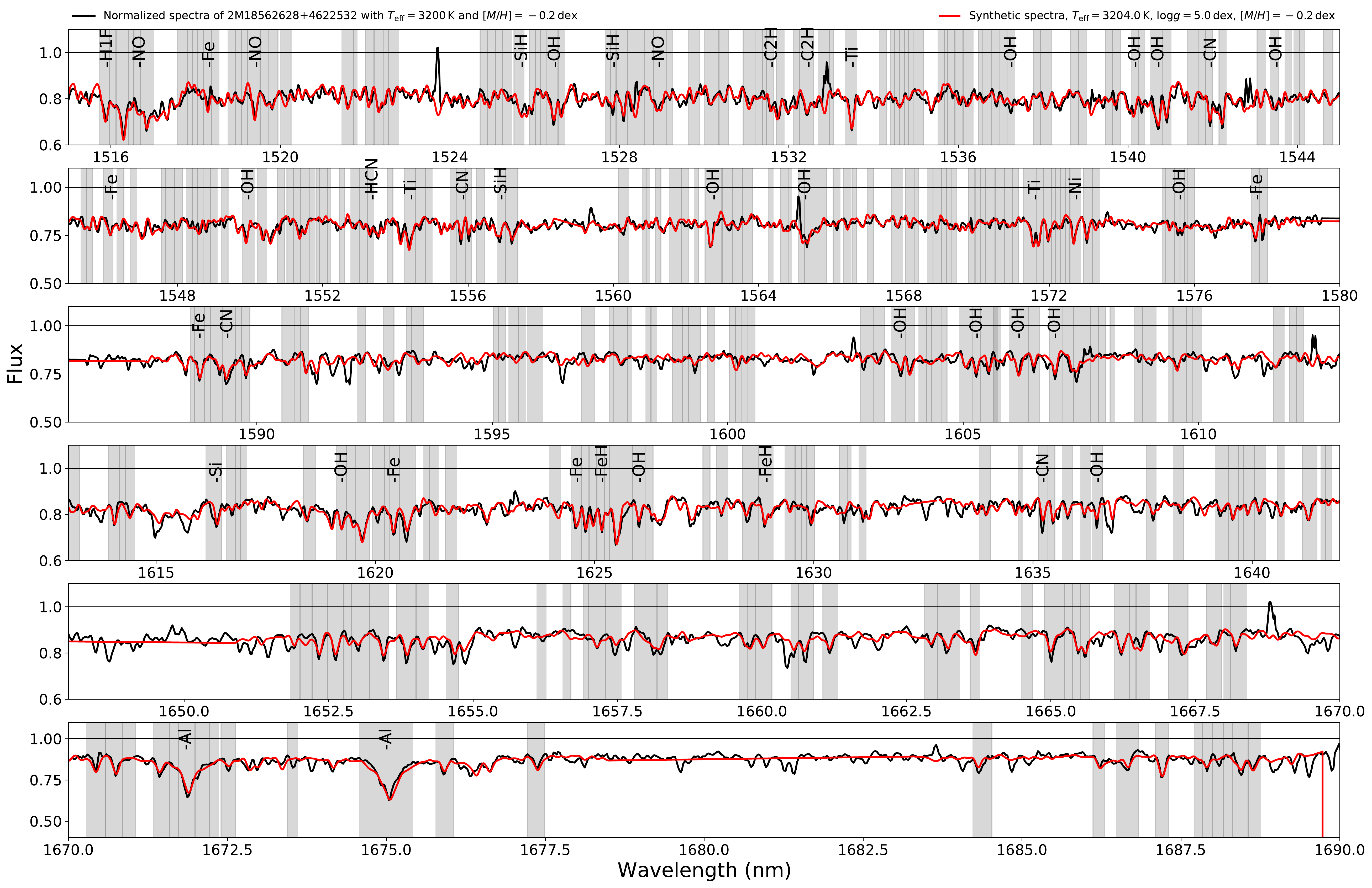}}
\caption{Comparison between APOGEE spectra of the star 2M18562628+4622532 (black, normalized using a synthetic spectra with $T_\mathrm{eff} = 3200\,$K, $\log g = 5.0$\,dex and $[M/H]=-0.2\,$dex) and the best fitting synthetic spectra (red, straight line) for APOGEE wavelength range.
In gray highlight are the areas used for $\chi ^2 $ minimization by our pipeline's algorithm.
The best fitting parameters derived were $T_\mathrm{eff}=3204\pm100$\,K,$\log g=4.99\pm0.2$\,dex, $[M/H] = -0.19\pm0.1$\,dex. The available ASPCAP parameters are $T_\mathrm{eff}=3430\pm58$\,K, $\log g=5.09\pm0.13$\,dex and $[M/H] = -0.44\pm0.02$\,dex.}
\label{Mdwarf_2}
\end{figure}

\end{landscape}

\begin{landscape}

\begin{figure}
\resizebox{\hsize}{!}{\includegraphics{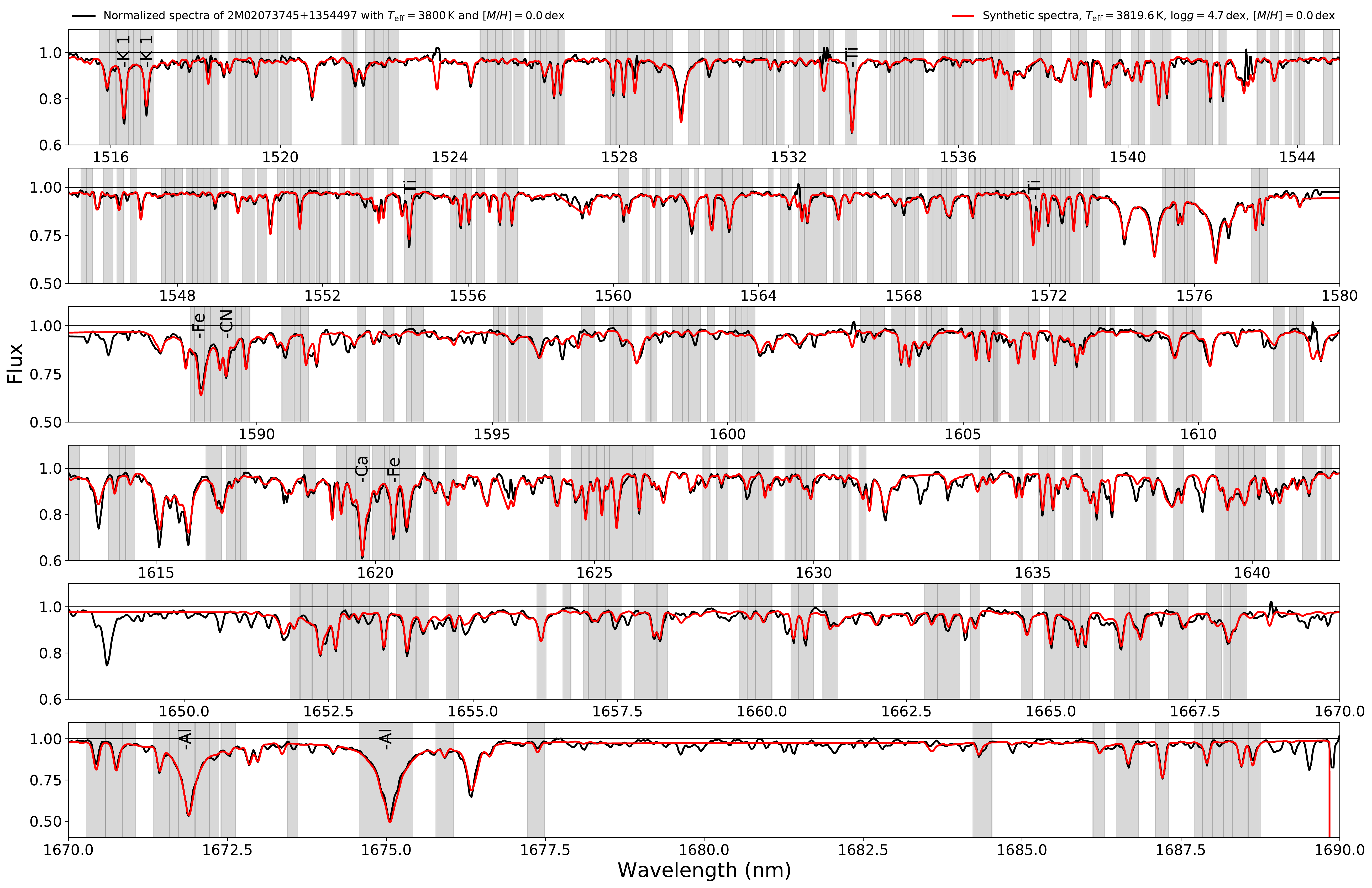}}
\caption{Comparison between APOGEE spectra of the star 2M02073745+1354497 (black, normalized using a synthetic spectra with $T_\mathrm{eff} = 3800\,$K, $\log g = 4.7$\,dex and $[M/H] = +0.0\,$dex) and the best fitting synthetic spectra (red, straight line) for APOGEE wavelength range.
In gray highlight are the areas used for $\chi ^2 $ minimization by our pipeline's algorithm.
The best fitting parameters derived were $T_\mathrm{eff}=3819\pm100$\,K,$\log g=4.69\pm0.2$\,dex, $[M/H] = +0.01\pm0.1$\,dex. The available ASPCAP parameters are $T_\mathrm{eff}=3866\pm63$\,K, $\log g=4.83\pm0.11$\,dex and $[M/H] = +0.11\pm0.01$\,dex.}
\label{Mdwarf_5}
\end{figure}

\end{landscape}

\begin{landscape}

\begin{figure}
\resizebox{\hsize}{!}{\includegraphics{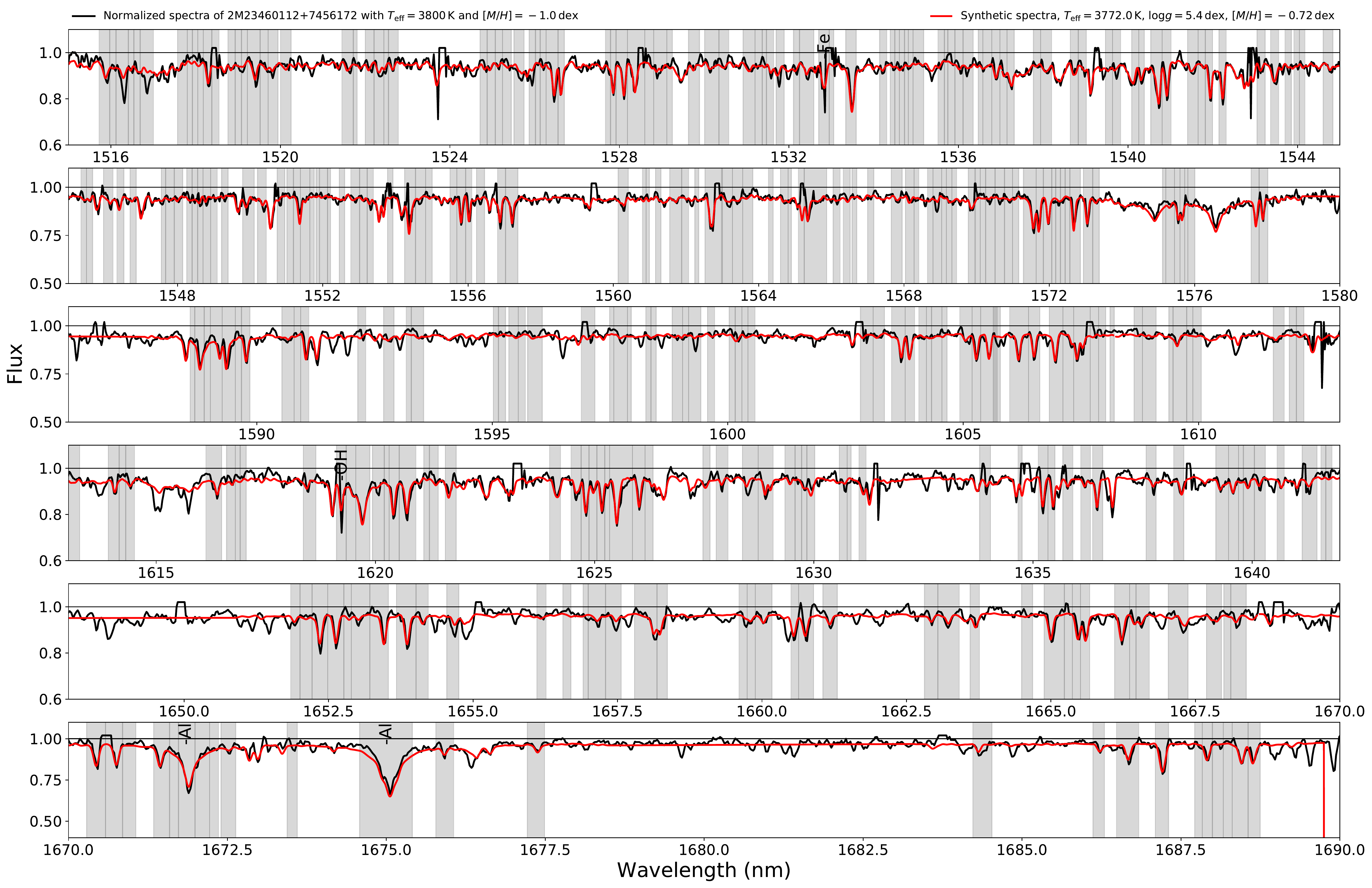}}
\caption{Comparison between APOGEE spectra of the star 2M23460112+7456172 (black, normalized using a synthetic spectra with $T_\mathrm{eff} = 3600\,$K, $\log g = 5.0$\,dex and $[M/H] = -0.6\,$dex) and the best fitting synthetic spectra (red, straight line) for APOGEE wavelength range.
In gray highlight are the areas used for $\chi ^2 $ minimization by our pipeline's algorithm.
The best fitting parameters derived were $T_\mathrm{eff}=3588\pm100$\,K,$\log g=5.13\pm0.2$\,dex, $[M/H] = -0.49\pm0.1$\,dex. The available ASPCAP parameters are $T_\mathrm{eff}=3560\pm62$\,K, $\log g=4.86\pm0.13$\,dex and $[M/H] = -0.44\pm0.02$\,dex.}
\label{Mdwarf_6}
\end{figure}

\end{landscape}

\begin{landscape}

\begin{figure}
\resizebox{\hsize}{!}{\includegraphics{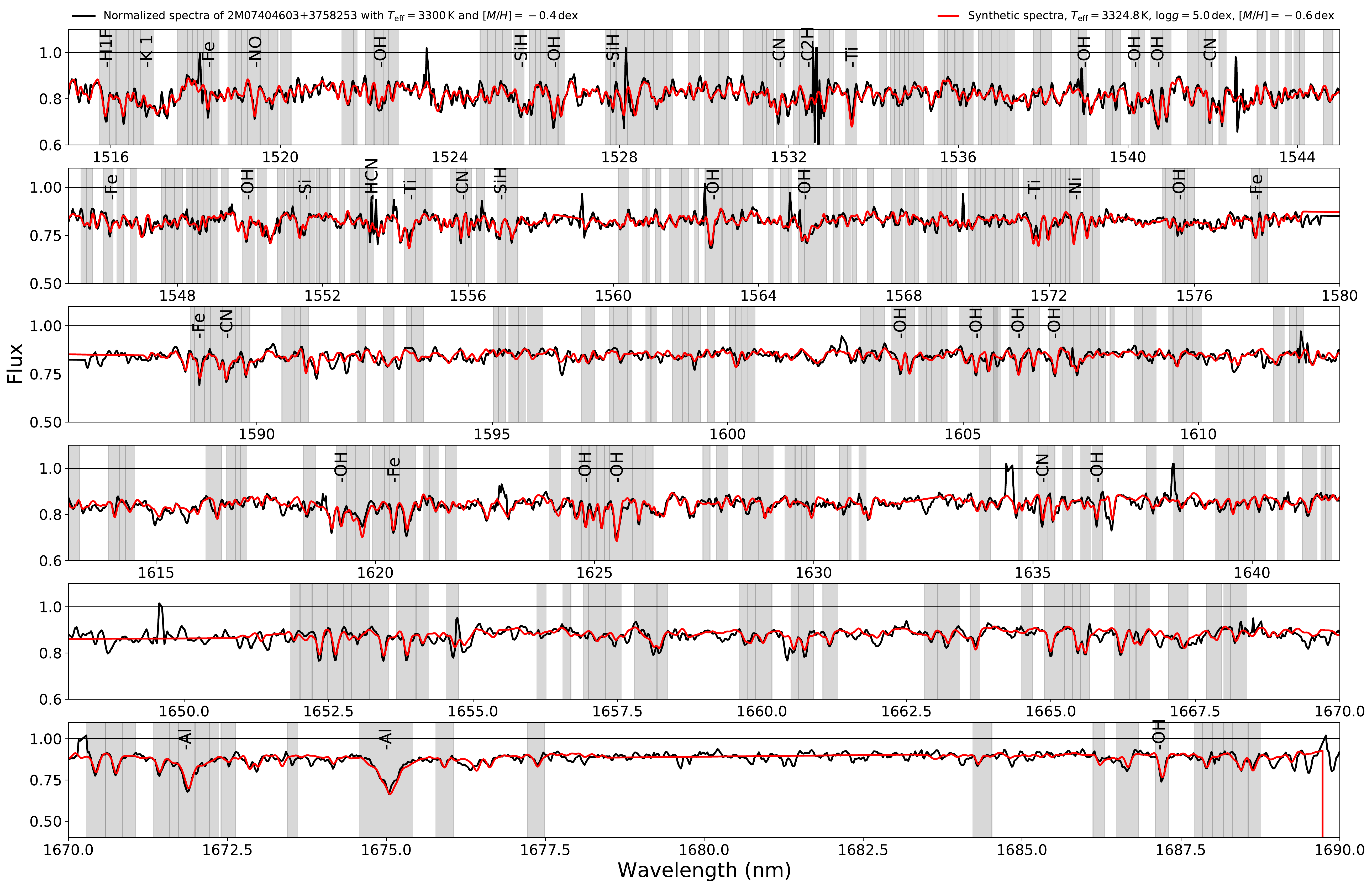}}
\caption{Comparison between APOGEE spectra of the star 2M07404603+3758253 (black, normalized using a synthetic spectra with $T_\mathrm{eff} = 3300\,$K, $\log g = 5.0$\,dex and $[M/H] = -0.4\,$dex) and the best fitting synthetic spectra (red, straight line) for APOGEE wavelength range.
In gray highlight are the areas used for $\chi ^2 $ minimization by our pipeline's algorithm.
The best fitting parameters derived were $T_\mathrm{eff}=3325\pm100$\,K,$\log g=4.97\pm0.2$\,dex, $[M/H] = -0.60\pm0.1$\,dex.}
\label{Mdwarf_7}
\end{figure}

\end{landscape}

\onecolumn


\section{Full results for stellar sample, DR16}

\begin{longtable}{lccccccc|ccc}
 \caption{\label{table:MdwarfResultsDR16}All derived parameters for stellar sample from APOGEE DR16. Uncertainties in derived parameters are 100\,K for $T_\mathrm{eff}$, 0.2\,dex for $\log g$, and 0.1\,dex for $[M/H]$. iSpec output parameters indicated by letter O. Best fitting normalization parameters indicated by letter N.}\\
 \hline\hline
{} & \multicolumn{7}{c}{iSpec Output} & \multicolumn{3}{c}{Normalization} \\
APOGEE\_ID & $T_\mathrm{eff}$ & $\log g $ & $\textrm{[M/H]}$ & v$_\mathrm{mic}$ & v$_\mathrm{mac}$ & ${v \sin i}$ & $\chi ^2$ & $T_\mathrm{eff}$ & $\log g $ & $\textrm{[M/H]}$ \\
\hline
\endfirsthead
\caption{continued.}\\
\hline\hline
{} & \multicolumn{7}{c}{iSpec Output} & \multicolumn{3}{c}{Normalization} \\
APOGEE\_ID & $T_\mathrm{eff}$ & $\log g $ & $\textrm{[M/H]}$ & v$_\mathrm{mic}$ & v$_\mathrm{mac}$ & ${v \sin i}$ & $\chi ^2$ & $T_\mathrm{eff}$ & $\log g $ & $\textrm{[M/H]}$ \\
\hline
\endhead
\hline
\endfoot

\hline
2M00004701+1624101	&	3787	&	4.77	&	0.02	&	0.00	&	5.83	&	5.95	&	0.04	&	3800	&	4.7	&	0	\\
2M00182256+4401222	&	3695	&	4.96	&	-0.54	&	0.73	&	6.21	&	4.31	&	0.03	&	3700	&	4.9	&	-0.6	\\
2M00204772+5936173	&	3604	&	4.79	&	-0.01	&	0.76	&	6.51	&	1.63	&	0.04	&	3600	&	4.8	&	0	\\
2M00243855+5119224	&	3506	&	4.83	&	-0.33	&	0.69	&	6.91	&	4.30	&	0.03	&	3500	&	5	&	-0.4	\\
2M00244054+5511590	&	3800	&	4.80	&	-0.40	&	0.83	&	5.79	&	0.00	&	0.03	&	3800	&	4.8	&	-0.4	\\
2M00251060+6724193	&	3715	&	4.78	&	-0.29	&	0.88	&	6.09	&	0.00	&	0.03	&	3700	&	4.8	&	-0.2	\\
2M00262872+6747026	&	3786	&	4.66	&	-0.06	&	0.05	&	5.81	&	4.38	&	0.03	&	3800	&	4.7	&	0	\\
2M00274401+5330504	&	3407	&	4.94	&	0.06	&	0.87	&	7.37	&	0.28	&	0.05	&	3400	&	4.8	&	0	\\
\end{longtable}

\section{Full results for stellar sample, DR14}

\begin{longtable}{lccccccc|ccc}
 \caption{\label{table:MdwarfResultsDR14}All derived parameters for stellar sample from APOGEE DR14. Uncertainties in derived parameters are 100\,K for $T_\mathrm{eff}$, 0.2\,dex for $\log g$, and 0.1\,dex for $[M/H]$. iSpec output parameters indicated by letter O. Best fitting normalization parameters indicated by letter N.}\\
 \hline\hline
{} & \multicolumn{7}{c}{iSpec Output} & \multicolumn{3}{c}{Normalization} \\
APOGEE\_ID & $T_\mathrm{eff}$ & $\log g $ & $\textrm{[M/H]}$ & v$_\mathrm{mic}$ & v$_\mathrm{mac}$ & ${v \sin i}$ & $\chi ^2$ & $T_\mathrm{eff}$ & $\log g $ & $\textrm{[M/H]}$ \\
\hline
\endfirsthead
\caption{continued.}\\
\hline\hline
{} & \multicolumn{7}{c}{iSpec Output} & \multicolumn{3}{c}{Normalization} \\
APOGEE\_ID & $T_\mathrm{eff}$ & $\log g $ & $\textrm{[M/H]}$ & v$_\mathrm{mic}$ & v$_\mathrm{mac}$ & ${v \sin i}$ & $\chi ^2$ & $T_\mathrm{eff}$ & $\log g $ & $\textrm{[M/H]}$ \\
\hline
\endhead
\hline
\endfoot

\hline

2M00004701+1624101	&	3790	&	4.77	&	0	&	0	&	5.82	&	5.78	&	0.16	&	3800	&	4.7	&	0	\\
2M00182256+4401222	&	3692	&	4.99	&	-0.59	&	0.8	&	6.24	&	4.04	&	0.03	&	3700	&	4.9	&	-0.6	\\
2M00182256+4401222	&	3702	&	4.98	&	-0.56	&	0.8	&	6.2	&	2.54	&	0.03	&	3700	&	4.9	&	-0.6	\\
2M00204772+5936173	&	3583	&	4.72	&	-0.01	&	0.61	&	6.58	&	4.03	&	0.21	&	3600	&	4.8	&	0	\\
2M00243855+5119224	&	3671	&	5.11	&	-0.54	&	0.83	&	6.37	&	4.4	&	0.04	&	3700	&	5	&	-0.8	\\
2M00244054+5511590	&	3793	&	4.6	&	-0.5	&	0.92	&	5.78	&	6.09	&	0.04	&	3800	&	4.8	&	-0.4	\\
2M00251060+6724193	&	3702	&	4.81	&	-0.31	&	1.09	&	6.14	&	6.17	&	0.06	&	3700	&	4.8	&	-0.2	\\
2M00262872+6747026	&	3844	&	4.8	&	-0.12	&	0.56	&	5.64	&	5.75	&	0.18	&	3900	&	4.7	&	-0.2	\\
\end{longtable}

\npnoround
 
\end{appendix}

\end{document}